\documentclass[preprint,12pt]{elsarticle}
\usepackage{amsmath,amssymb,hyperref}
\usepackage[dvips]{epsfig}
\usepackage{graphicx}
\usepackage{xcolor}

\usepackage{multirow}
\usepackage{siunitx} 
\usepackage{bbm}
 \usepackage{booktabs} 
\PassOptionsToPackage{unicode}{hyperref}
\PassOptionsToPackage{naturalnames}{hyperref}

\hypersetup{unicode=false}

\definecolor{heidelbeer}{rgb}{0.5,0,0.5}

\usepackage{amssymb}
\usepackage[utf8x]{inputenc}
\usepackage{amsmath}
\DeclareMathOperator{\sech}{sech}
\usepackage{amsmath,bm}
\usepackage{lipsum}

\usepackage{graphicx}
\usepackage{bbm}
\usepackage{booktabs}
\usepackage{bm}
\usepackage{color,xcolor}
\usepackage{epsfig}

\usepackage{float}
\journal{Physics Letters B}

\begin{document}

\begin{frontmatter}

\title{Electron-positron pair creation induced by multi-pulse train of electric fields: effect of randomness in time-delay}

\author[1,2]{Deepak Sah}

\author[1,2]{Manoranjan P. Singh}

\address[1]{Theory and Simulations Lab,Theoretical and Computational Physics Section, Raja Ramanna Centre for Advanced Technology, Indore-452013, India}
\address[2]{Homi Bhabha National Institute, Training School Complex, Anushakti Nagar, Mumbai 400094, India}

\begin{abstract}
We investigate the creation of electron-positron pairs (EPPs) in a sequence of alternating-sign, time-dependent electric field pulse trains by solving the quantum Vlasov equations. Specifically, we focus on Sauter-like pulse trains with random time delays between successive pulses, drawn from a Gaussian distribution wherein the extent of fluctuations is controlled by the standard deviation $\sigma_T$ of the distribution. We find that increasing $\sigma_T$ leads to a dramatic transformation in the longitudinal momentum spectrum. The well-known fringe pattern, akin to that in the multi-slit interference, gets significantly modified. The averaged spectra exhibit a robust Gaussian-like envelope with residual oscillations, which are much more prominent in the central momentum region. 
Notably, we find that in certain cases, stochastic time delays lead to a pronounced enhancement in the central peak of the distribution function for pulse train containing $N$ pulses.
For example, for $N=20$ pulses, $\sigma_T \approx 31$ $[m^{-1}]$(about $17\%$
of the mean time delay) yields nearly a tenfold increase in the central peak, which for $\sigma_T \approx 50$ $[m^{-1}]$ (about $27\%$ of the mean time delay), scales up to $10^3.$
This may open up new possibilities for optimizing multi-pulse field configurations and guide future experimental designs aimed at maximizing EPPs creation.


\end{abstract}

\end{frontmatter}

\section{Introduction}
\label{introduction}
The spontaneous creation of electron-positron pairs(EPPs) from vacuum in the presence of intense external fields is a fundamental prediction of quantum electrodynamics (QED) \cite{Sauter:1931zz, Heisenberg:1936nmg}. However, observing this effect experimentally remains challenging due to the significant exponential suppression, given by $\exp(-\pi E_c/E)$, where $E_{\text{c}} = m^2/|e| \approx 1.3 \times 10^{16} \text{V/cm}$ represents the Schwinger critical field strength, $m$ is the electron mass, $e$ is the electron charge (the units $\hbar =c=1$ are used) and $E$ is the applied field strength \cite{Schwinger:1951nm}. The laser intensity needed to reach this threshold is approximately $I_{\text{c}} \approx 10^{29}~\text{W/cm}^2$, which greatly surpasses the capabilities of the current conventional laboratory systems. Nonetheless, significant progress in high-intensity laser technology and the construction of cutting-edge laser facilities \cite{DiPiazza:2011tq, Mourou:2014kea, Ringwald:2001ib, Bell:2008zzb}  is steadily closing the gap, making the experimental observation of this phenomenon increasingly feasible. This progress continues to inspire extensive theoretical and experimental efforts. Currently, laser systems have reached peak intensities of approximately $10^{23}~\text{W/cm}^2$ \cite{Yoon:2021ony}.
\par
EPPs creation can occur through various mechanisms under strong electromagnetic fields. One example is the Bethe-Heitler mechanism, in which a super-intense laser interacts with the Coulomb field of a nucleus, resulting in pair creation \cite{Augustin:2014xga,Muller:2003zz}. Another widely studied mechanism is the Breit-Wheeler mechanism, where a high-energy gamma photon collides with an ultra-strong laser field to produce pairs \cite{DiPiazza:2016maj,Krajewska:2012eb}.
Notably, the only direct experimental observation of positron production via such mechanisms was carried out at the Stanford Linear Accelerator Center (SLAC) \cite{Burke:1997ew}, where a $46.6$~GeV electron beam was made to interact with a terawatt laser pulse of intensity around $10^{18}~\text{W/cm}^2$. In that setup, positrons were generated following nonlinear Compton scattering, which produced photons that then triggered the Breit-Wheeler process. Apart from the nonlinear Breit-Wheeler process, other strong-field QED effects—such as nonlinear Compton scattering \cite{Ilderton:2020dhs} and strong-field-induced vacuum pair production—have also attracted substantial attention.
\par
To investigate pair creation in different external field configurations, researchers have developed various theoretical approaches. These studies primarily focus on reducing the required field strength and enhancing the yield of produced pairs \cite{Kohlfurst:2017git,Fedotov:2022ely, Dumlu:2010vv, Hebenstreit:2011wk}.   Semiclassical techniques such as the generalized Wentzel-Kramers-Brillouin (WKB) approximation \cite{Brezin:1970xf, Kim:2003qp} and the worldline instanton method \cite{Kim:2000un,Dunne:2005sx,Schubert:2011fz} have been widely used to describe pair production probabilities. Quantum kinetic approaches, including the quantum Vlasov equation (QVE)\cite{Sah:2025nbr,Hebenstreit:2014lra,Banerjee:2018azr}, the low-density approximation \cite{Abdukerim:2015dsa,Blaschke:2012vf}, and the Dirac-Heisenberg-Wigner (DHW) formalism \cite{Bialynicki-Birula:1991jwl,Hebenstreit:2010vz,Li:2025paq}, offer more detailed quantum descriptions. Brezin and Itzykson \cite{Brezin:1970xf} analyzed pair production in a time-dependent, spatially homogeneous electric field using the WKB approximation, deriving probabilities based on the Keldysh adiabaticity parameter, $\gamma = m\omega/|e|E$, which determines the interaction regime, with $\omega$ being the frequency of the electric field. 
\par
Over the past decade, investigations have shown that the momentum spectrum of created particles is highly sensitive to the profile of the applied electric field and its parameters, particularly in the tunneling regime \cite{Hebenstreit:2009km}. This sensitivity extends to the multiphoton and the intermediate regimes too \cite{Banerjee:2018azr,Sah:2025nbd}. Furthermore, recent studies indicate that these dependencies significantly influence the time evolution of the momentum spectrum as well; see \cite{Sah:2023jlz,Sah:2024qcs} for details. 

A major advancement in the study of vacuum pair creation has been the realization that structured multi-pulse electric field configurations can dramatically enhance pair production due to quantum interference effects. Analogous to the optical double-slit experiment, time-domain multiple-slit interference has emerged as a key mechanism that both increases the total yield and shapes the features of the momentum spectrum of the created pairs. Akkermans and Dunne \cite{Akkermans:2011yn} were the first to demonstrate Ramsey-type multiple-time-slit interference, showing that in an alternating-sign $N$-pulse electric field, the central peak of the momentum distribution scales as $N^2$, indicating constructive interference. Building on this concept, Kohlfürst \cite{Kohlfurst:2012yw} explored various multi-pulse configurations, illustrating how precise pulse shaping can be used to optimize the pair production rate. In addition, combinations of different pulse types have been shown to enhance the pair creation. For instance, the dynamically assisted Schwinger mechanism, which involves the interplay of a strong, slowly varying pulse with a weak, rapidly oscillating one, has been demonstrated to significantly boost pair production \cite{Schutzhold:2008pz,Olugh:2019nej}. Among the other field configurations, multi-pulse electric field comprising of sequences of time-dependent pulses with alternating signs have garnered particular interest. In such setups, not only do parameters of individual pulses, such as amplitude, duration, and shape, influence EPPs production, but the temporal spacing between pulses also plays a decisive role. Prior studies~\cite{Hebenstreit:2014lra,Jansen:2016crq,Granz:2019sxb,Ilderton:2019ceq,Kaminski:2018ywj} have shown that the inter-pulse time delay can significantly affect the momentum distribution of the produced pairs. Notably, Ref.~\cite{Granz:2019sxb} reports that the total pair production probability exhibits damped oscillations as a function of the time interval between pulses. 

\par
The aforesaid theoretical models assume pulse trains with a uniform inter-pulse delay.   However, real-world experimental conditions may deviate from this idealization. Fluctuations in pulse timing can arise due to limitations in synchronization and control, particularly in sequences involving a large number of pulses subject to the shot-to-shot variations \cite{Yi:2025cpr,Saldin:2002zs}.  One may also consider experiments wherein a pulse train is derived from multiple sources which may not be well synchronized. This raises a key question as to how do random fluctuations in pulse timing influence EPPs production.  Specifically, does the breakdown of perfect coherence diminish the enhancements typically observed in the usual case of uniform inter-pulse delay, or can certain stochastic realizations of random temporal spacing in the multi-pulse train unexpectedly enhance pair creation? What happens to the momentum spectrum? How different it is when averaged over the many realizations of the random inter-pulse delays from that for the single realization. To address these questions, we investigate vacuum pair production under field configurations with random time delays between successive pulses, which are drawn from a Gaussian distribution wherein the degree of randomness is quantified by the standard deviation \( \sigma_T \)  of the distribution. We set $\mu_T = 180.32\ [m^{-1}]$ to facilitate direct comparison with earlier studies of regular pulse trains~\cite{Li:2014xga}, but we also examine the role of $\mu_T$ in the stochastic regime. While $\mu_T$ influences individual realizations, we show in the Supplementary Material that the ensemble-averaged momentum spectrum is robust to small variations in $\mu_T$. The key parameter governing the transition from coherent to incoherent spectra is $\sigma_T$.





It is found that increasing the value of \( \sigma_T \) has a strong influence on the longitudinal momentum spectrum. The well-known fringe-like structure in the spectrum of the usual non-stochastic counterpart arising due to the interference effect intrinsic to strong-field QED (for example, in Ref. \cite{Akkermans:2011yn}) gets modified.
The momentum spectrum, when averaged over the realizations of the time delays with a given \( \sigma_T \),  exhibits a robust Gaussian-like envelope with residual oscillations which are much more pronounced in the central momentum region.  Intriguingly, sufficiently large values of $\sigma_T$ lead to a pronounced enhancement of the central peak in the momentum spectrum. For instance, in the case of $N=20$, a value of $\sigma_T \approx 31~[m^{-1}]$ yields nearly a tenfold increase, while around $\sigma_T \approx 50~[m^{-1}]$ the enhancement reaches almost three orders of magnitude.
\par
This paper is organized as follows. In Sec.~\ref{theory}, we introduce the theoretical formalism based on the quantum Vlasov equation. In Sec.~\ref{result}, we present and discuss our numerical results. In Sec.~\ref{summary}, we provide a brief conclusion and outlook.
\section{Theoretical Framework}
\label{theory}

In the subcritical field regime, $E < E_{\text{c}}$, pair production and the corresponding back-reaction current are minimal. This allows us to neglect both collision effects and the internal electric field. Moreover, the spatial focusing scale of typical laser pulses is usually much larger than the electron Compton wavelength—which characterizes the spatial extent relevant for vacuum pair creation—so spatial inhomogeneities of the background field are strongly suppressed. 
Nevertheless, we emphasize that spatial variations and accompanying magnetic fields can, in general, have a significant impact on electron–positron pair production. This has been demonstrated in several works, including 
Refs.~\cite{Aleksandrov:2016lxd,Kohlfurst:2015niu,Ruf:2008ahs,Hebenstreit:2010vz,Kohlfurst:2017git,Kohlfurst:2021skr}.
\newline
In the present work we restrict ourselves to a simplified but widely used scenario: a spatially uniform, purely time-dependent electric field. When two counterpropagating laser pulses form a standing wave with sufficiently large beam waist, the associated magnetic field near the antinodes can be neglected, effectively giving $B(t) = 0$. Consequently, the background field can be modeled as $\bm{E}(t) = (0, 0, E(t))$.

Adopting the temporal gauge $A^0(t) = 0$, the corresponding four potential is given by $A^\mu(t) = (0, 0, 0, A(t))$,  where $A(t)$ is related to the electric field  as $E(t) = -\dot{A}(t)$. Pair creation from the vacuum in the presence of such an external electric field has been studied using the quantum Vlasov equation (QVE) by many researchers; see, for example, \cite{Banerjee:2018azr,Abdukerim:2013vsa,Otto:2014ssa} and references therein. QVE is a standard tool within the framework of quantum kinetic theory, and its detailed derivation is readily available in the literature, e.g., in Ref. \cite{Schmidt:1998vi}. Here, for the sake of completeness, we provide only the essential equations and the notations.
\par
Starting from the Dirac equation in a homogeneous electric field and employing a time-dependent Bogoliubov transformation, the QVE can be formulated as an integro-differential equation that governs the evolution of the single-particle momentum distribution function $f(\bm{p}, t)$:
\begin{equation}
\frac{d f(\bm{p}, t)}{dt} = \frac{\lambda(\bm{p},t)}{2}  \int_{t_0}^{t} dt' 
\lambda(\bm{p}, t')[1 - 2 f(\bm{p}, t')] 
\cos\!\left[\Theta (\bm{p}, t, t')\right] ,
\label{eqn1}
\end{equation}
where $ \lambda (\bm{p},t) = \frac{e E(t) \varepsilon_{\perp}}{\omega^2(\bm{p}, t)}$, is the amplitude of the vacuum transition, while $\Theta (\bm{p},t,t') = 2 \int_{t'}^{t} d\tau \, \omega(\bm{p}, \tau )$
stands for the dynamical phase, describing the vacuum oscillations modulated by the external field. 
The quasiparticle  energy  $\omega(\bm{p}, t),$ the transverse energy $\varepsilon_{\perp}$ and longitudinal quasiparticle  momentum $P_3$ are defined as:
\begin{align}
     \omega(\bm{p}, t) &= \sqrt{\varepsilon_{\perp}^2 + P_3^2(p_3,t)},
     \label{eqn2}
\end{align}
\begin{align}
     \varepsilon_{\perp} &= \sqrt{m^2 + p_{\perp}^2},
     \label{eqn3}
\end{align}
\begin{align}
    P_3(t) &= p_3 - eA(t),
    \label{eqn4}
\end{align}
where $\bm{p} = (p_{\perp}, p_3)$ represents the canonical momentum. Here, $p_{\perp} =|p_{\perp}| =\sqrt{p_1^2 + p_2^2}$ is the modulus of the momentum component perpendicular to the electric field, and $p_3$ stands for the momentum component parallel to the electric field $E(t).$  
\par
It is important to note that the distribution function $f(\bm{p},t)$ represents the number of real particles created with momentum $\bm{p}$ in the asymptotic limit $t \to +\infty$. This limit corresponds to a physical scenario in which the external laser field vanishes. Our analysis focuses on the distribution function in the asymptotic regime. 

\par
 Eq.~\eqref{eqn1} is difficult to solve numerically due to the presence of rapidly oscillating phase term in the integrand. 
 It is, therefore, convenient to recast this equation as an equivalent system of three coupled ordinary differential equations:
\begin{equation}\label{ODE1}
 \frac{df(\bm{p},t)}{dt}=\frac{1}{2} \lambda(\bm{p},t)u(\bm{p},t),
\end{equation}
\begin{equation}\label{ODE2}
 \frac{du(\bm{p},t)}{dt}=\lambda(\bm{p},t)[1-2f({\bm{p}},t)]-2\omega(\bm{p},t)v(\bm{p},t),
\end{equation}
\begin{equation}\label{ODE3}
 \frac{dv(\bm{p},t)}{dt}=2\omega(\bm{p},t)u(\bm{p},t).
\end{equation}
Together with the initial conditions
$f(\bm{p},-\infty)=u(\bm{p},-\infty)=
v(\bm{p},-\infty)=0$
this set of equations becomes a well-defined and numerically  solvable initial
value problem.

\begin{figure}[tbp]\suppressfloats
\centering
\includegraphics[width=0.9682\columnwidth]{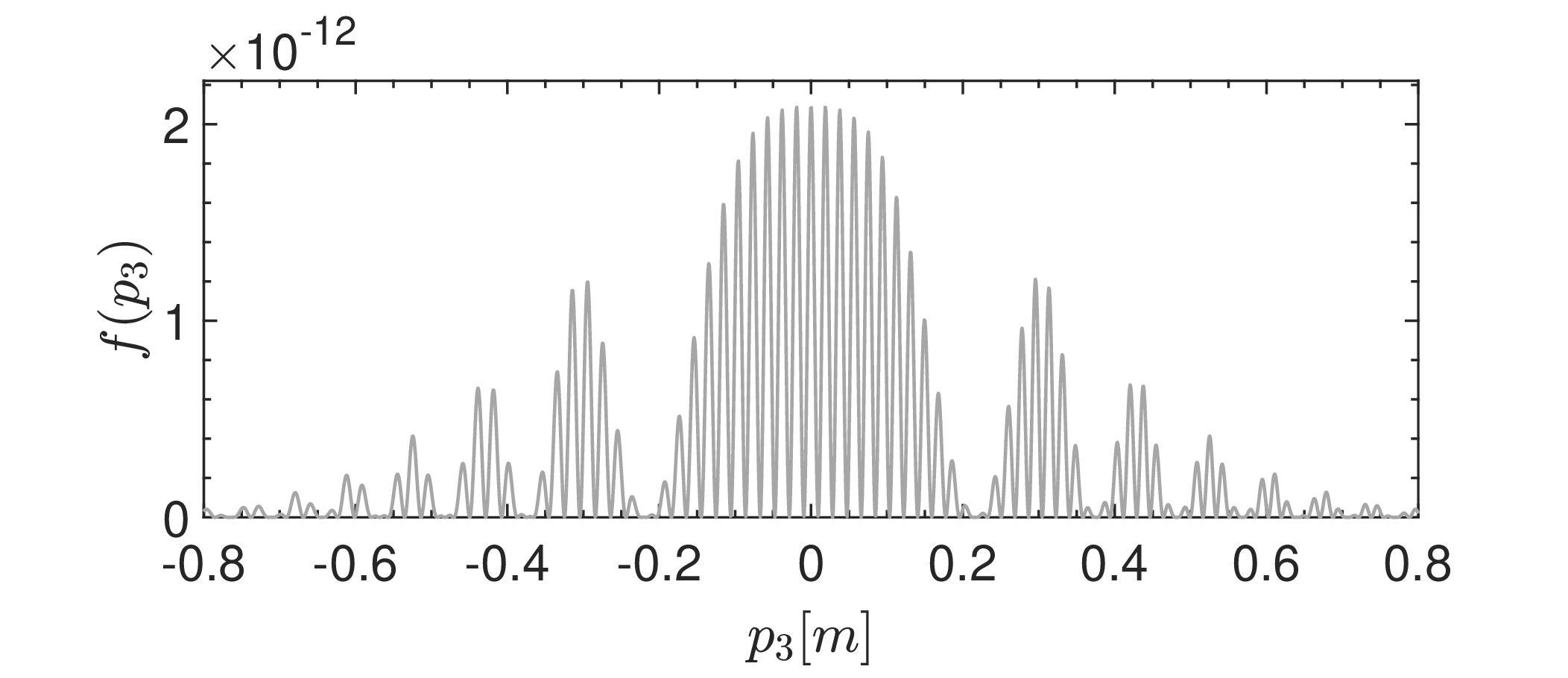}
\caption{\label{Fig0} The longitudinal momentum spectrum of created EPPs in an alternating-sign four-pulse electric field $E(t)$ for the non-stochastic case ($\sigma_T = 0$). The transverse momentum is set to zero, and all quantities are expressed in units of the electron mass. The field parameters are $E_0 = 0.1E_c$, $\tau = 20~[m^{-1}]$, and $\mu_T = 180.32~[m^{-1}]$.}
\end{figure}

\section{Results}
\label{result}
We consider a field configuration composed of a sequence of alternating-sign, time-dependent Sauter electric pulses referred to as a multi-pulse train: 
\begin{align}   \label{field}
    E(t) &= \sum_{k=1}^{N} (-1)^{k-1} E_0 \, \sech^2 \left( \frac{t + \left(k - \frac{N+1}{2} \right) T_k}{\tau} \right),
    \end{align}
where $E_0$ denotes the amplitude of each electric field pulse, and $N$ is the total number of pulses in the pulse train.  
The random variable $T_k$ specifies the temporal position of the $k$th pulse, whose center is located at $(\frac{(N+1-2k)}{2})T_k$.
Thus, the timing of pulses in the pulse train becomes stochastic. This electric field should be compared with those in Refs.~\cite{Li:2014xga,Akkermans:2011yn}   where all the pulses in the pulse train are regularly spaced with a fixed time delay. Henceforth, we shall refer to such a pulse train as regular or non-stochastic. The electric field in Eq.~\eqref{field} reduces to the one considered in Refs.~\cite{Akkermans:2011yn,Li:2014psw,Kohlfurst:2012yw} upon replacing the random variables $\{T_k\}$  by a constant which is the fixed time delay between successive pulses.  Note that the time delay between the $i$th and the $j$th pulse for the electric field considered here (Eq. \eqref{field}) depends on the random variables $T_i$ and $T_j$.
The corresponding vector potential is given by,
\begin{align}   \label{potential}
    A(t) &=    - E_0 \tau  \Biggl[  1 + \sum_{k=1}^{N} (-1)^{k-1}  \tanh{\left( \frac{t + \left(k - \frac{N+1}{2} \right) T_k}{\tau} \right)} \Biggr].
  \end{align}

We take random variables $\{T_k\}$ to be independent and identically distributed (IID) according to a Gaussian distribution having a mean  $\mu_{T}$ and variance $\sigma_{T}^2 $.   Specifically, we generate $\{T_k\}$  using MATLAB’s built-in \texttt{randn} function~\cite{MATLAB:randn}, which returns an array of pseudo-random numbers drawn from the standard normal distribution with zero mean and unit variance (i.e., unbounded and symmetric about zero).  
Accordingly, 
\begin{align}
\{T_k\} &= \mu_{T} + \sigma_{T} \times \text{randn}(1,N),
\label{Tdelay}
\end{align}
It is evident from Eq.~\eqref{Tdelay} that the stochastic pulse train with random variables $\{T_k\}$ would turn into a non-stochastic one, with $\mu_T$  being the delay between the successive pulses, when $\sigma_T$ is set to zero.  In other words,  for the stochastic pulse train, the fluctuation in the time delay about its mean $\mu_T$ is governed by the standard deviation $\sigma_T$.  We fix the mean as $\mu_T = 180.32 [m^{-1}]$, which is exactly the same as that considered for the non-stochastic pulse train in Ref.~\cite{Li:2014xga}.


\begin{figure}[tbp]
\centering
{\includegraphics[width=0.9682\columnwidth]{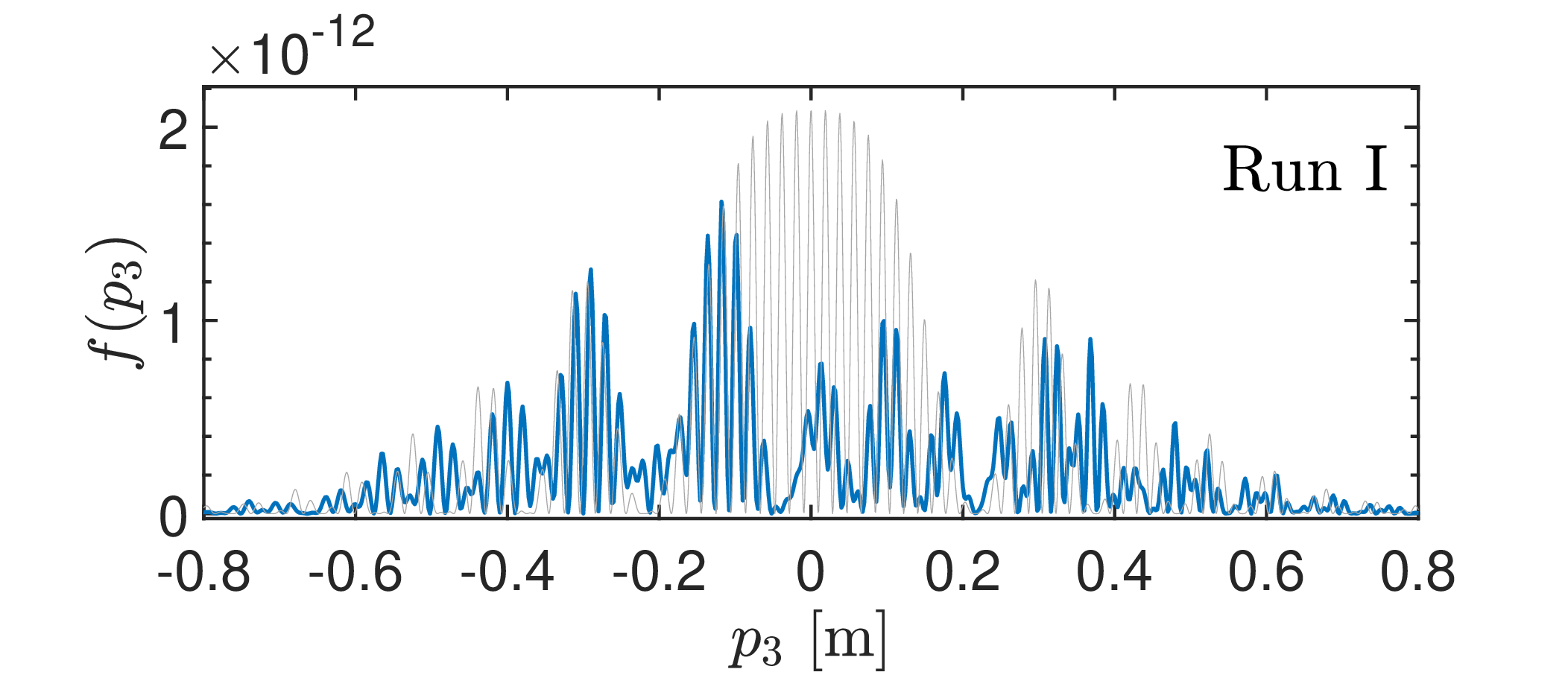}
\includegraphics[width=0.9682\columnwidth]{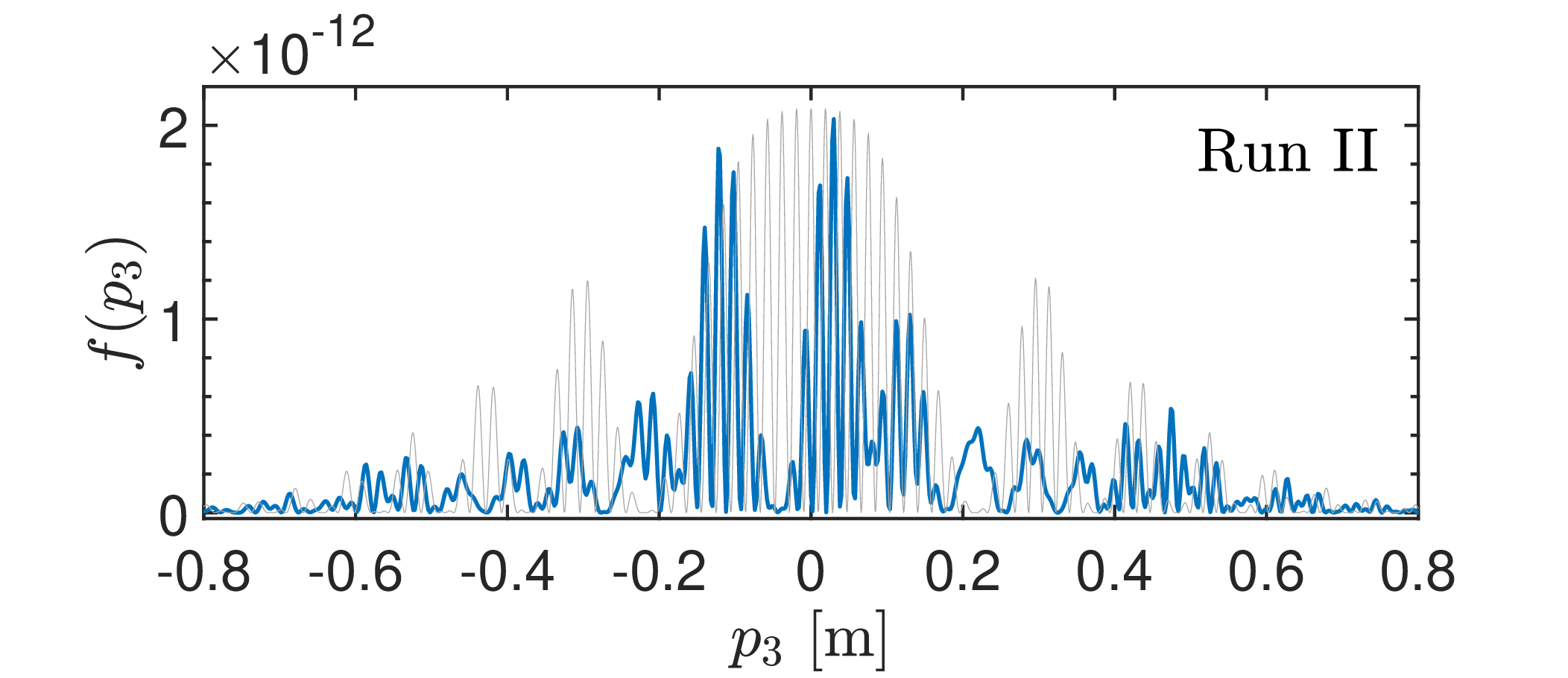}
\includegraphics[width=0.9682\columnwidth]{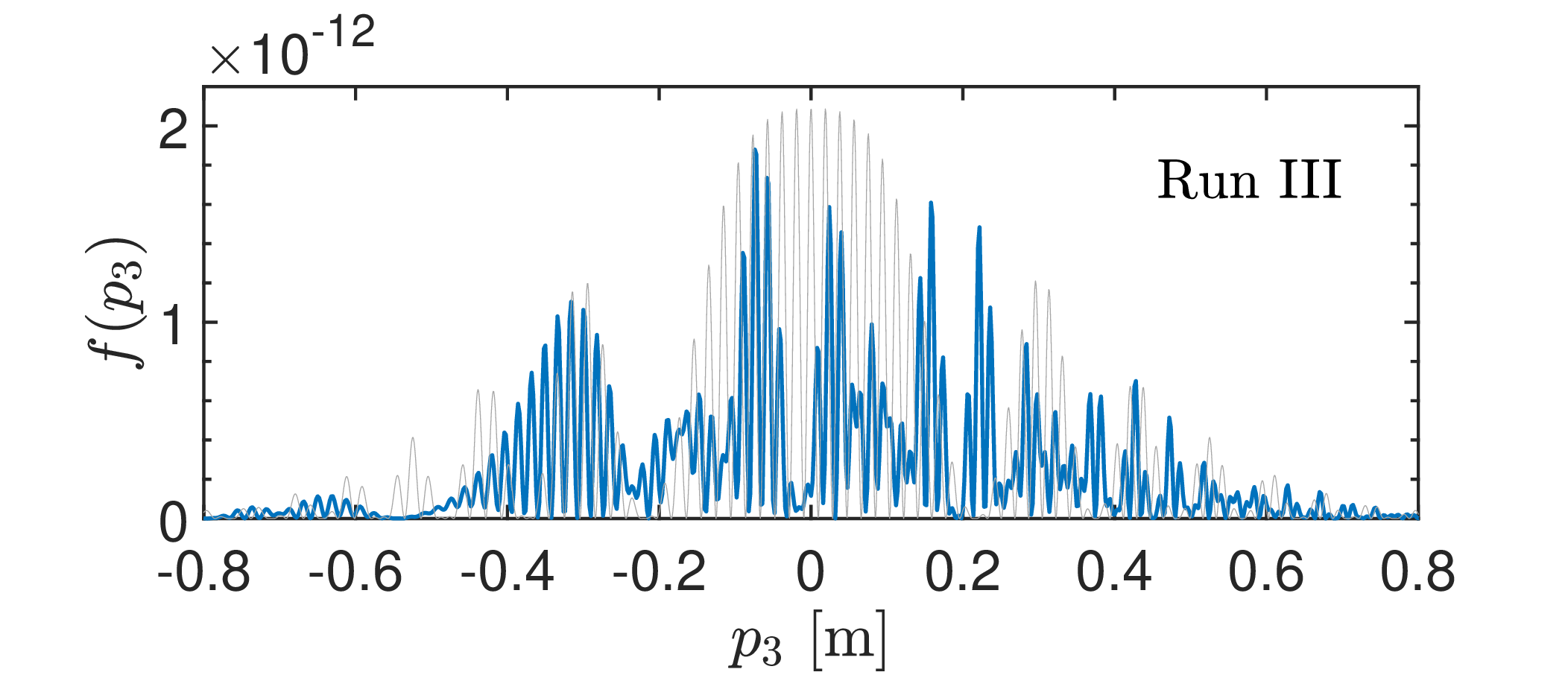}
}
\caption{\label{Fig7} The longitudinal momentum spectrum of created EPPs in an alternating-sign $N$-pulse electric field $E(t)$ with $N=4$. The blue line shows a stochastic realization with $\sigma_T = 15 \,[m^{-1}]$, while the grey line represents the non-stochastic case. The three panels correspond to independent realizations (Runs I–III). The value of transverse momentum is taken to be zero, and all the units are taken in electron mass units. The electric field parameters are $E_0 = 0.1 E_c,$ $\tau = 20[m^{-1}],$ and $\mu_T = 180.32[m^{-1}]$. }
\end{figure}

To demonstrate the effect of randomness in the stochastic pulse train on the longitudinal momentum spectrum of the created pairs, we numerically solve the system of first-order ordinary differential equations given in Eqs.~\eqref{ODE1}-\eqref{ODE3},  We plot the resulting momentum spectra for both the stochastic and non-stochastic cases. The latter,  which is widely studied in the literature, serves as a reference to identify the modifications introduced by the randomness. 
In Figure~\ref{Fig0}, the longitudinal momentum spectrum of created EPPs is shown for the non-stochastic case ($\sigma_T = 0$) of an alternating-sign $N$-pulse electric field $E(t)$. The spectrum exhibits a characteristic $N$-slit interference fringe-like structure consisting of several bands of maxima and minima. Most pairs are contained in the central band located around $p_3 \approx 0$, which has many peaks of nearly equal heights exhibiting the well-known $N^2$ scaling ~\cite{Akkermans:2011yn,Li:2014xga}.  Furthermore, the central band has a gradually varying and broad momentum profile. On the other hand, the successive side bands, located symmetrically on either side of the central band, have far fewer peaks. The farther the side band from the central band, the lower the peak height of the maxima therein. Therefore, with increasing value of $|p_3|$, the fringes gradually vanish asymptotically.  Overall, the spectrum exhibits a highly regular, symmetric interference pattern, with the central broad band ($-0.2[m] < p_3 < 0.2 [m]$) dominating the distribution.

In contrast to the non-stochastic case, Figure~\ref{Fig7} shows the spectra for the stochastic case ($\sigma_T=15[m^{-1}]$) with $N=4$ for three independent realizations (Runs I–III). 
Introducing randomness modifies the regular $N$-slit interference pattern observed in the non-stochastic case.
The spectra lose their symmetry about $p_3=0$, and the central band becomes distorted in a realization-dependent manner. Although a broad Gaussian-like envelope persists, the peaks fragment into irregular structures with uneven heights and spacings. The differences between Runs I–III highlight run-to-run fluctuations, reflecting the sensitivity of the momentum distribution to stochastic variations in the time delays. The corresponding values of the random variable $\{T_k\}$ are listed in Table~\ref{tab:T_all}. In Run I, the interference structure around the central region ($-0.2 [m] < p_3 < 0.2 [m] $) is strongly modified compared to the non-stochastic case. The sharp, well-ordered fringe pattern with evenly spaced maxima and minima is replaced by irregular sub-bands. The central peak at $p_3 = 0$ is noticeably suppressed, and the minima between fringes no longer drop to zero, reducing overall fringe visibility. On the left side ($-0.2 [m] < p_3 < 0 [m]$), relatively strong oscillations remain, with amplitudes comparable to those in the deterministic case. In contrast, on the right side ($0 [m] < p_3 < 0.2[m]$), the fringes become weaker and more irregular. Side-band peaks at $p_3 \approx -0.4[m]$ and $-0.3[m]$ remain visible, whereas their positive-$p_3$ counterparts are distorted and less pronounced. In Run II, the central interference band fragments into fewer sub-bands than in Run I, but the peak heights remain comparable to the non-stochastic case (grey curve). The spectrum is still dominated by the central maximum at $p_3 \approx 0$, much like in the deterministic case. Some side-band structures survive, but their peaks are noticeably suppressed in height and lose their regular spacing. As a result, the distribution becomes asymmetric about $p_3 = 0$, with the side-bands appearing less sharp and more irregular compared to the non-stochastic spectrum. In Run III, the asymmetry is most pronounced. One side of the spectrum (negative $p_3$) exhibits stronger and denser peaks, while the other side (positive $p_3$) is highly irregular and suppressed. The central band is significantly distorted, with one side dominating the distribution. This extreme imbalance demonstrates the strong sensitivity of the momentum spectrum to random variations in inter-pulse delays.
Overall, randomness in the inter-pulse delays ($\sigma_T = 15 [m^{-1}]$) modifies the highly regular fringe pattern of the non-stochastic case. The randomness not only destroys the ordered fringe and side-band hierarchy but also induces a clear left–right asymmetry across all runs. Although band-like interference features persist, their internal structure becomes irregular and asymmetric, with the degree of distortion varying from run to run. These run-to-run fluctuations highlight the stochastic nature of the driving field and its impact on the symmetry of the momentum distribution.
\begin{figure}[tbp]
\suppressfloats
\centering

{\includegraphics[width=0.9682\columnwidth]{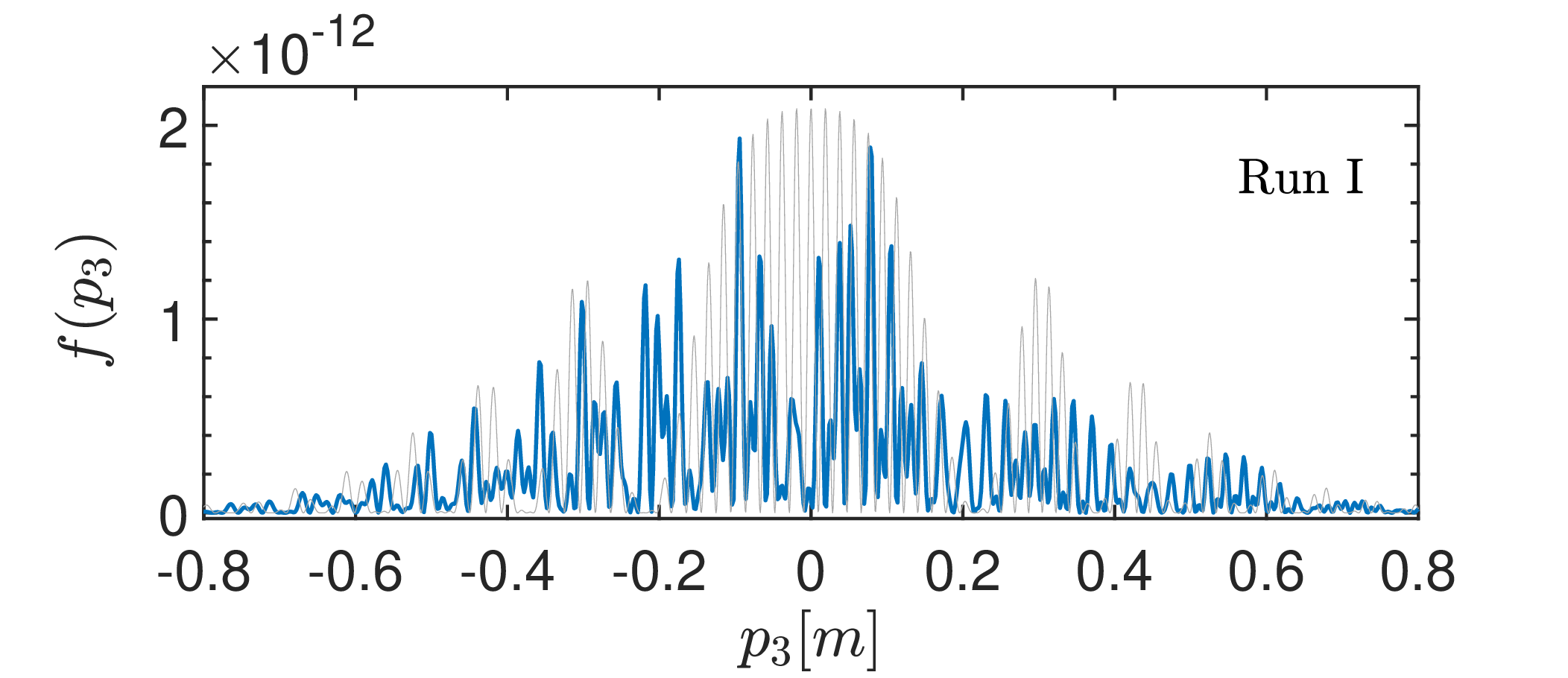}
\includegraphics[width=0.9682\columnwidth]{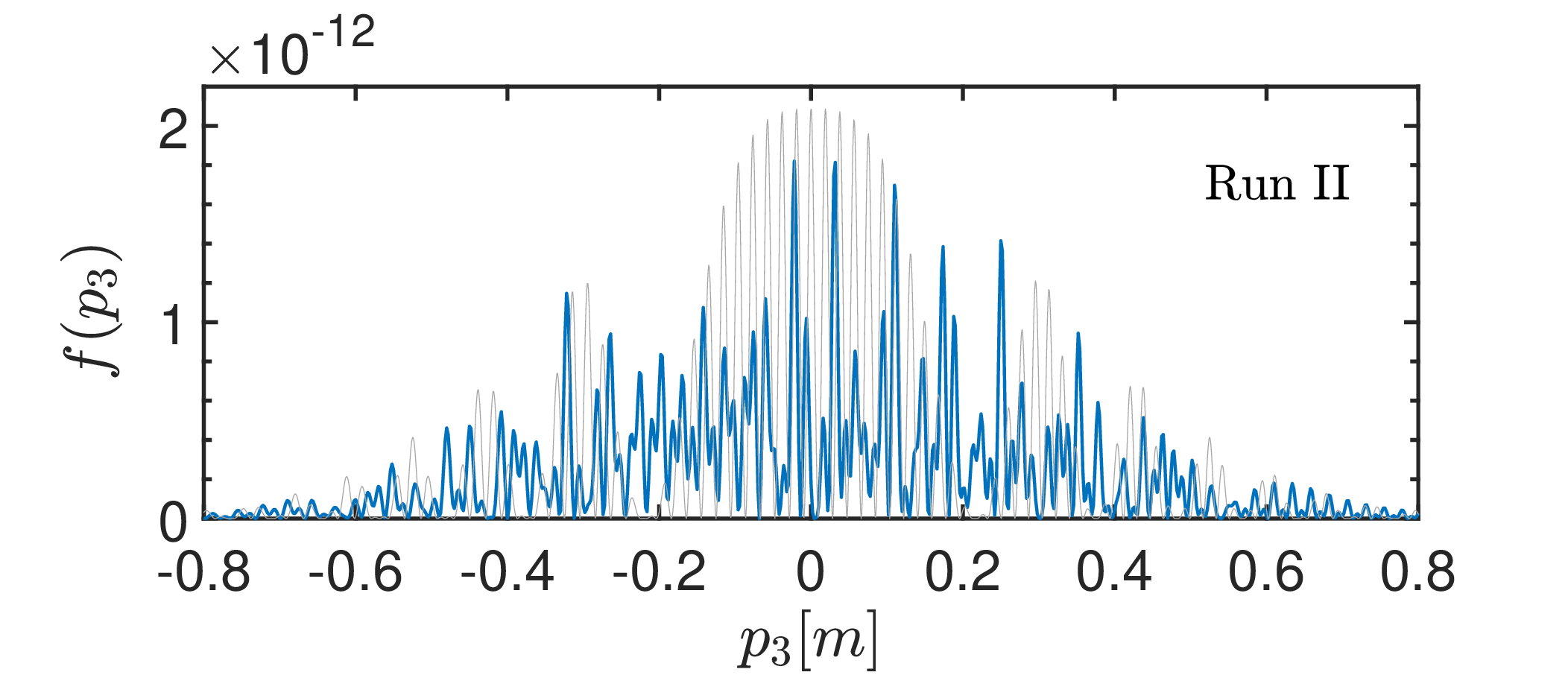}
\includegraphics[width=0.9682\columnwidth]{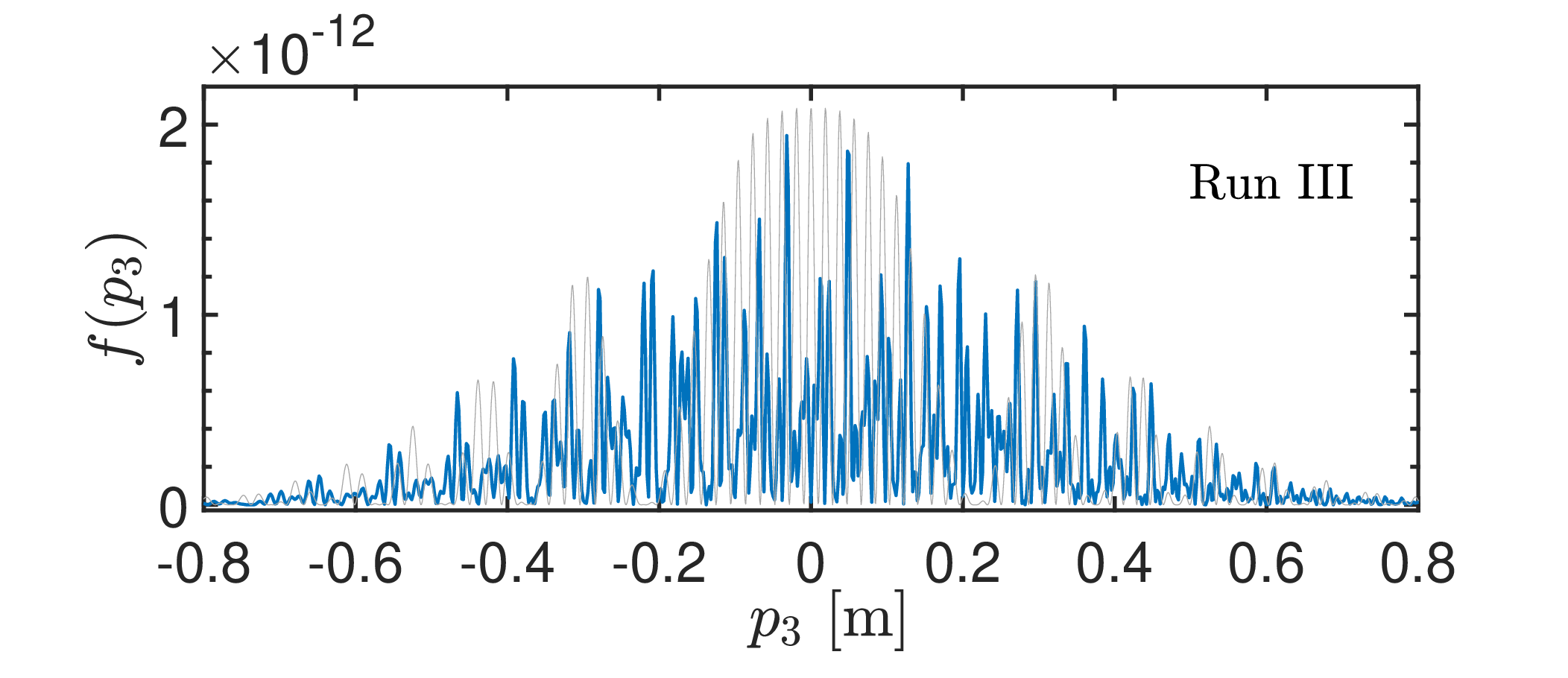}
}

\caption{\label{Fig8} Same as in Fig.~\ref{Fig7}, except for an alternating-sign four-pulse electric field with 
$\sigma_T = 45[m^{-1}]$.}
\end{figure}
\par
In Figure~\ref{Fig8}, the spread of randomly distributed inter-pulse delays ($\sigma_T = 45~[m^{-1}]$) further amplifies the stochasticity in the $N=4$ pulse trains, with the corresponding random variables ${T_k}$ tabulated in Table~\ref{tab:T_all}. Compared to the non-stochastic case and the weaker randomness ($\sigma_T = 15 [m^{-1}]$), the spectra exhibit fragmented peaks and reduced fringe visibility. In Run I, the central band ($-0.2[m] < p_3 < 0.2 [m]$) remains the most prominent feature; however, instead of exhibiting evenly spaced oscillations, it fragments into clusters of irregular peaks, while the side bands become distorted and lose symmetry about $p_3 = 0$, highlighting how stronger randomness enhances fragmentation and disrupts the interference hierarchy. Run II shows a pronounced left–right asymmetry: peaks for $p_3 < 0$ are enhanced, whereas those for $p_3 > 0$ are suppressed, and interference features are unevenly spaced, reflecting a breakdown of regular phase correlations. In Run III, the overall envelope resembles the deterministic case, yet the fine structure is highly disordered; individual peaks survive but no longer form a recognizable fringe pattern, and the asymmetry is weaker than in Run II. In general, increasing the inter-pulse delay spread to $\sigma_T=45 [m^{-1}]$ makes the momentum distribution highly sensitive to randomness: the spectra become fragmented, asymmetric, and run-dependent, demonstrating the strong stochastic influence of the driving field.

\begin{table}[htbp]
\centering
\begin{tabular}{|c|c|cc|cc|cc|}
\hline
$\sigma_T~[m^{-1}]$ & $k$ 
& \multicolumn{2}{c|}{Run I} 
& \multicolumn{2}{c|}{Run II} 
& \multicolumn{2}{c|}{Run III} \\
\hline
&  & $T_k$ & Pulse Centre & $T_k$ & Pulse Centre  & $T_k$ & Pulse Centre  \\
\hline
\multirow{4}{*}{15} 
& $1$ & 172.25 & 258.37 & 193.33 & 289.99 & 194.96 & 292.44 \\
& $2$ & 185.38 & 92.69  & 165.38 & 82.69  & 172.47 & 86.24  \\
& $3$ & 160.86 & -80.43 & 194.08 & -97.04 & 182.97 & -91.48 \\
& $4$ & 191.65 & -287.47& 171.45 & -257.17& 207.99 & -311.98\\
\hline
\multirow{4}{*}{45} 
& $1$ & 156.10 & 234.15 & 219.35 & 329.02 & 224.24 & 336.36 \\
& $2$ & 195.50 & 97.75  & 135.50 & 67.75  & 156.78 & 78.39  \\
& $3$ & 121.93 & -60.97 & 221.60 & -110.80& 188.27 & -94.13 \\
& $4$ & 214.30 & -321.45& 153.70 & -230.55& 263.32 & -394.98\\
\hline
\multirow{4}{*}{75} 
& $1$ & 139.95 & 209.93 & 245.36 & 368.04 & 253.51 & 380.28 \\
& $2$ & 205.62 & 102.81 & 105.61 & 52.81  & 141.07 & 70.54  \\
& $3$ & 83.01  & -41.51 & 249.12 & -124.56& 193.56 & -96.78 \\
& $4$ & 236.95 & -355.44& 135.96 & -203.94& 318.65 & -477.97\\
\hline
\end{tabular}
\caption{Tabulated values of the random variables $\{T_k\}$ [see Eq.~\ref{Tdelay}]  with $\mu_T = 180.32~[m^{-1}]$. Each block corresponds to independent realizations (Runs I–III). The last column in each run lists the corresponding pulse centers, located at $\tfrac{3}{2}T_1,\,\tfrac{1}{2}T_2,\,-\tfrac{1}{2}T_3,\, -\tfrac{3}{2}T_4$ for $N=4$.}
\label{tab:T_all}
\end{table}
\par
\begin{figure}[tbp]\suppressfloats
\centering
{\includegraphics[width=0.9682\columnwidth]{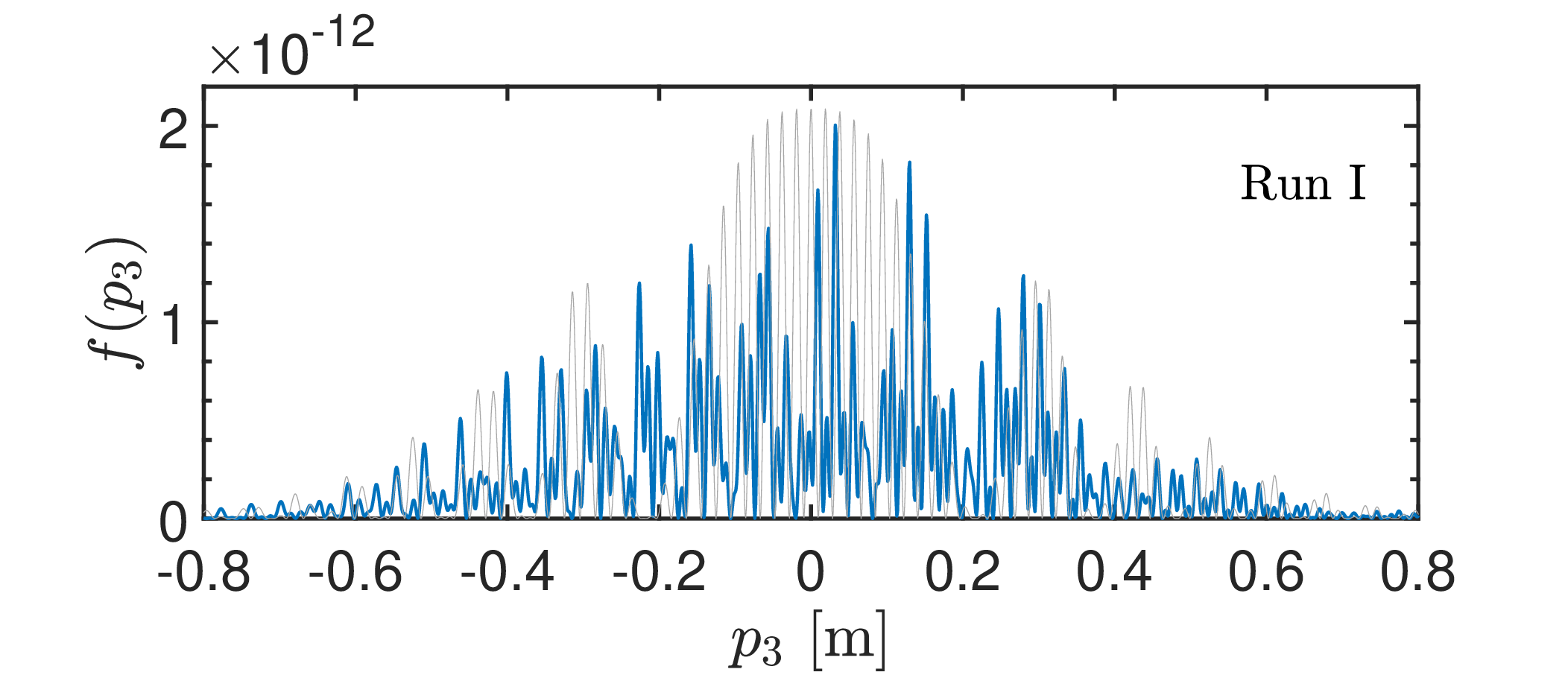}
\includegraphics[width=0.9682\columnwidth]{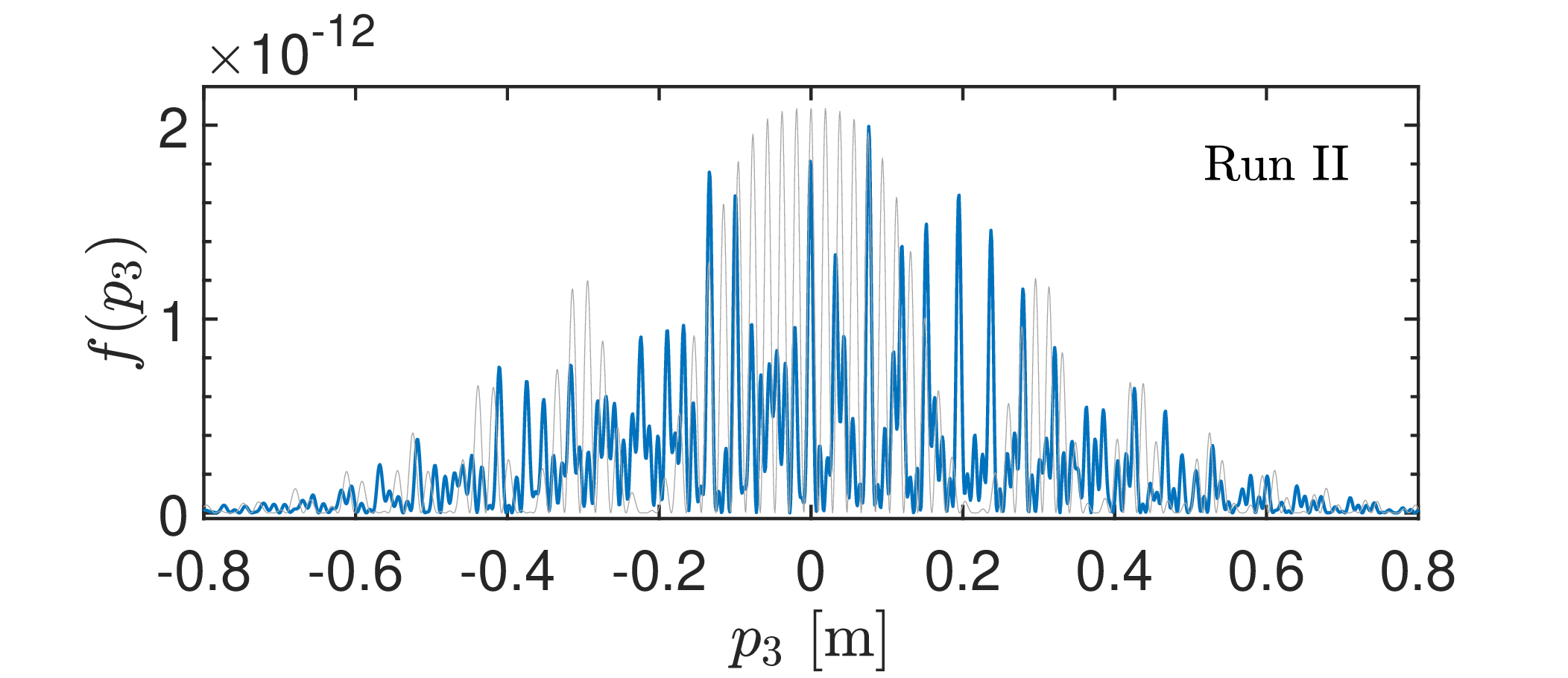}
\includegraphics[width=0.9682\columnwidth]{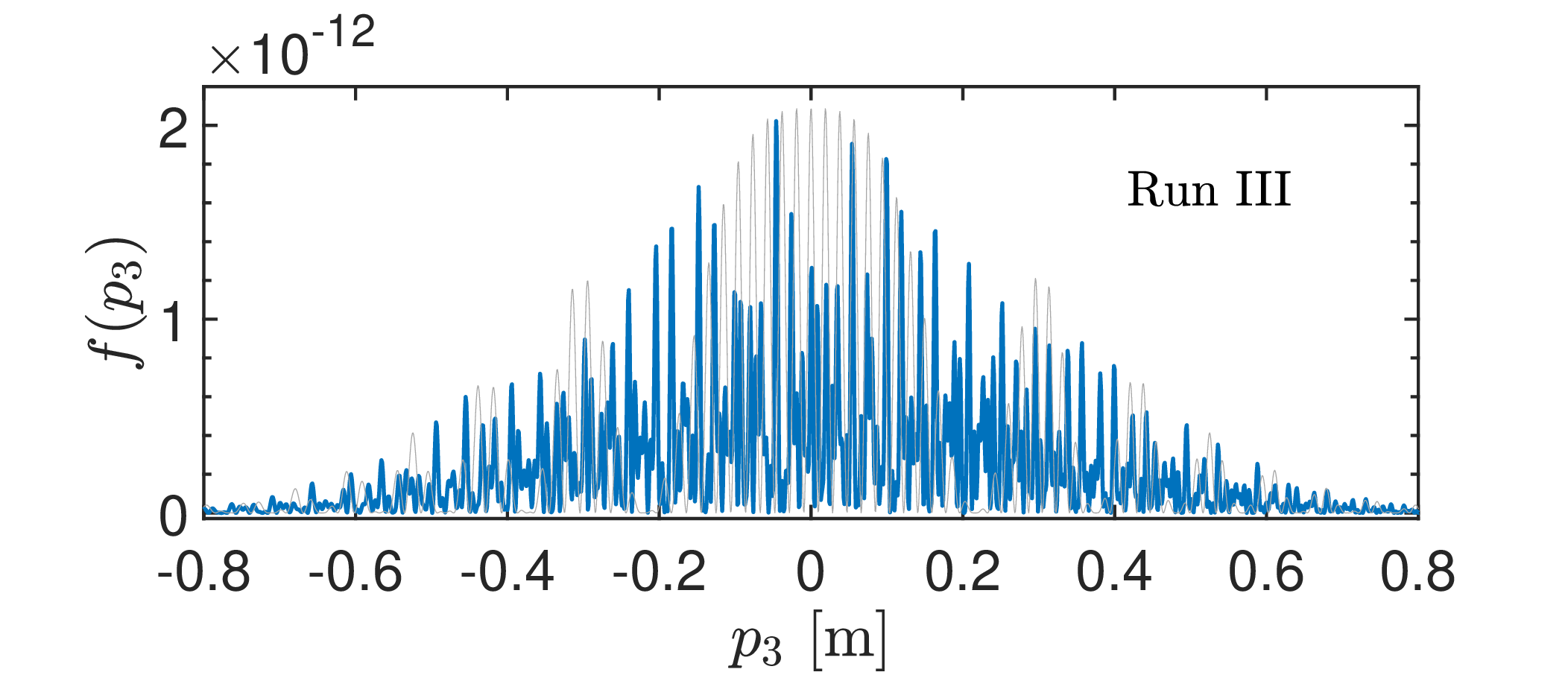}
}
\caption{\label{Fig9} Same as in Fig.~\ref{Fig7}, except for an alternating-sign four-pulse electric field with $\sigma_T = 75 \,[m^{-1}]$.}
\end{figure}

The spectra shown in Figure ~\ref{Fig9} correspond to the strongest degree of randomness considered, namely $\sigma_T = 75[m^{-1}]$. At this level of stochasticity, the longitudinal momentum spectra lose the well-separated bands of maxima and minima observed at smaller $\sigma_T$. Instead of regular clusters, the distributions exhibit isolated peaks scattered across a broad momentum range. Nevertheless, it is notable that the magnitude of the central peak remains comparable, or in some cases even slightly enhanced, compared to spectra at lower $\sigma_T$. In Run I, the spectrum exhibits a dense arrangement of sharp peaks centered around the central region ($p_3 \approx 0$), while the overall fringe-like band structure has disappeared. In Run II, the central peak is the most pronounced among the three realizations, with a strong maximum at $p_3 \approx 0$ accompanied by asymmetries more visible on one side of the spectrum. In Run III, the central region becomes more fragmented. Instead of a single dominant peak, multiple maxima of comparable height appear near $p_3=0$. This creates a visibly more irregular and distorted spectrum compared to Runs I and II. The side regions are highly suppressed, showing little evidence of structured oscillations. Taken together, the three realizations show that at $\sigma_T = 75[m^{-1}]$, the spectral structure becomes entirely run-dependent. Although all cases retain a concentration of spectral weight near the origin, the detailed arrangement of peaks varies significantly from run to run.
For $ N=4$ pulses, a comparative analysis of all figures reveals a distinct trend: as $\sigma_T$ increases, the spectral profile evolves from a well-defined interference pattern into a progressively incoherent and asymmetric distribution. The case $\sigma_T = 0$ represents the fully deterministic limit, serving as a reference for coherent particle production with a symmetric spectrum around the central momentum. Intermediate values ($\sigma_T = 15$ and $45[m^{-1}]$) show a gradual disappearance of the fringe-like bands of maxima and minima, accompanied by asymmetries in the peak heights and positions, indicative of increasing temporal randomness. At $\sigma_T = 75[m^{-1}]$, stochastic effects dominate, and the spectrum becomes strongly asymmetric, with the fringe-like features completely washed out.

\par
As discussed above, the longitudinal momentum spectra are highly sensitive to variations in inter-pulse delays. Increasing randomness progressively smears the well-defined fringe-like patterns of maxima and minima, suppressing several peaks and highlighting the effect of temporal disorder on coherent pair production.  To quantify the statistical impact of this stochasticity, we now analyze the averaged momentum spectrum, obtained by ensemble-averaging over many numerical runs with randomized time delays. In particular, we consider ensemble averages for two representative values of randomness, $\sigma_T = 15[m^{-1}]$ and $\sigma_T = 45[m^{-1}]$. Although these values do not span the entire parameter space, they are sufficient to capture the essential trends. Unlike earlier results, which reflected spectra from individual random realizations, ensemble averaging is crucial for extracting statistically robust and physically meaningful features. This approach also mirrors realistic experimental scenarios, where multiple experimental shots must be accumulated and averaged to obtain reliable signatures from stochastic sources.
\par
\begin{figure}[tbp]\suppressfloats
\centering
\includegraphics[width=0.980\columnwidth]{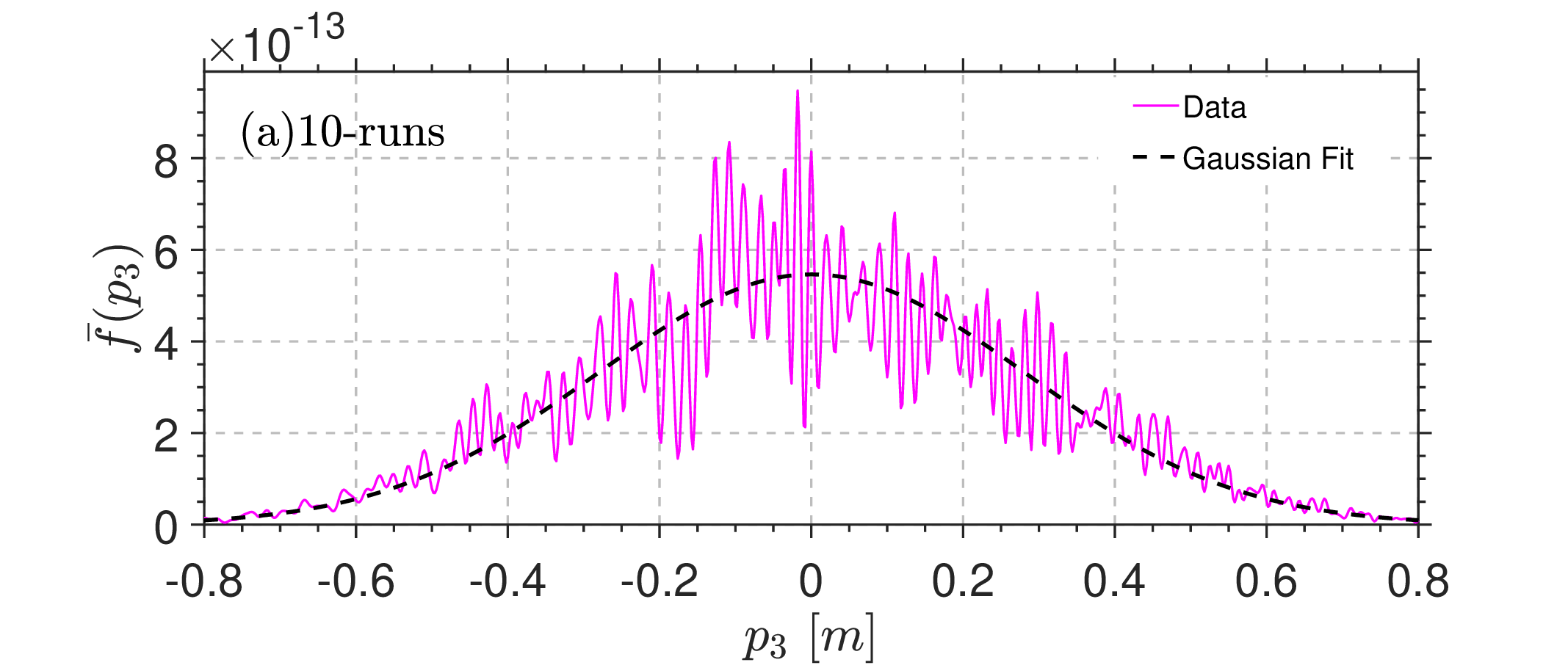}
\includegraphics[width=0.980\columnwidth]{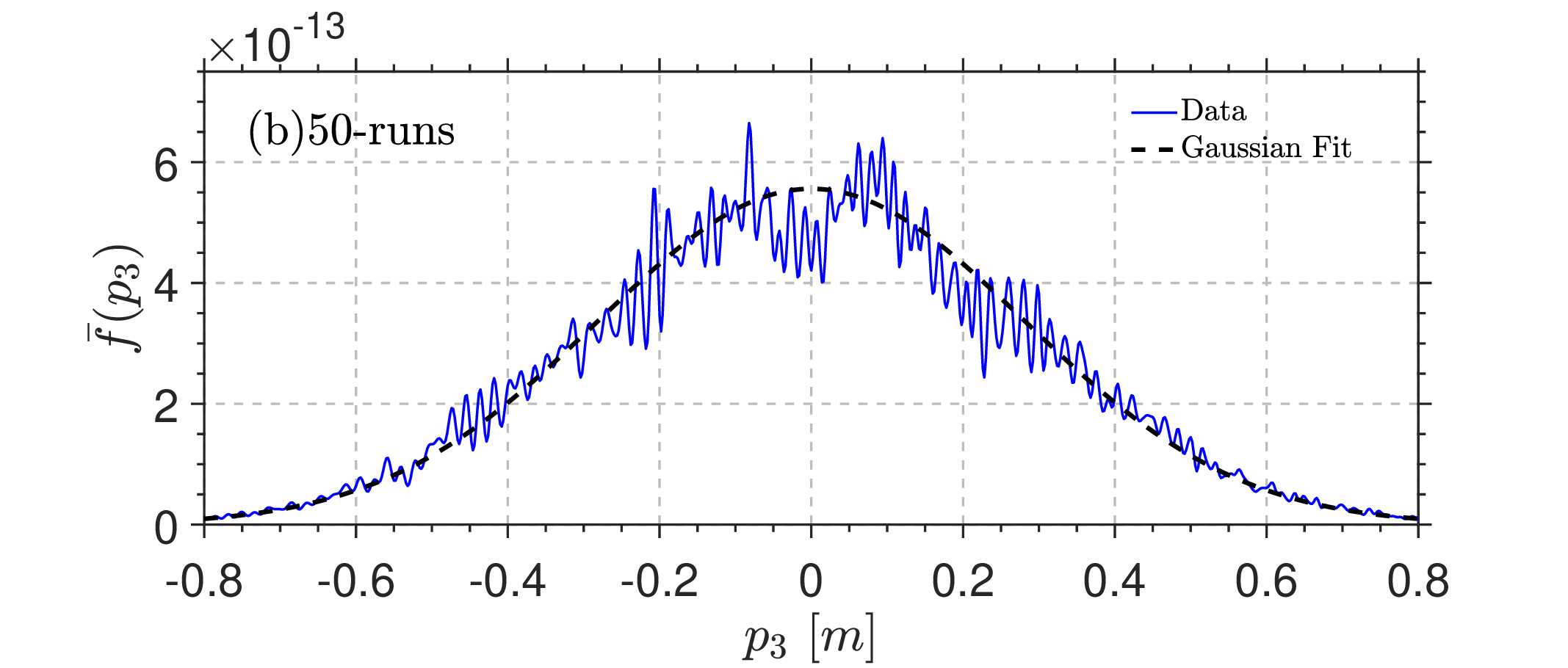}
\includegraphics[width=0.980\columnwidth]{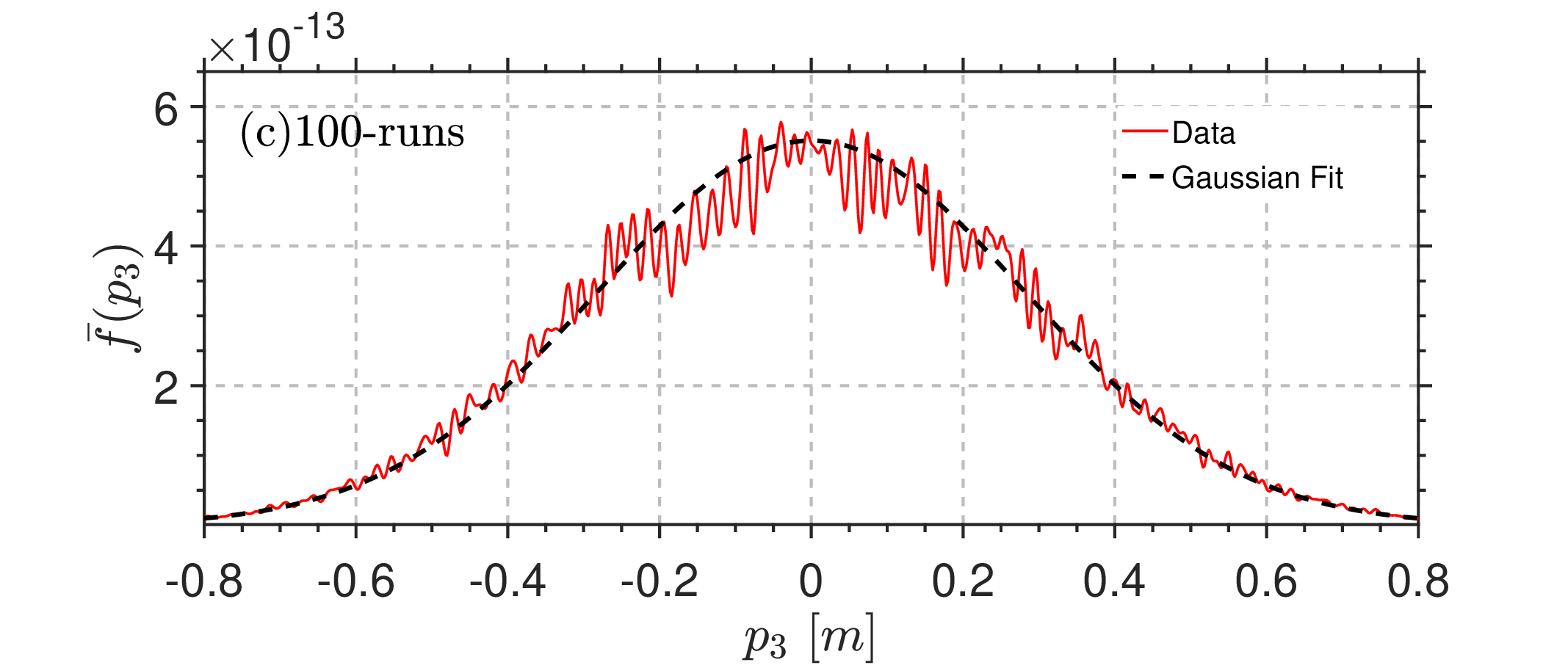}
\caption{\label{Fig9.1}
 Averaged momentum spectra $\bar{f}(p_3)$ computed over different numbers of random samples with randomized time delays for an alternating-sign four-pulse electric field $E(t)$. Each panel corresponds to a different number of averaging runs, with the black dashed curve indicating a Gaussian fit. The field parameters are $E_0 = 0.1 E_c$, $\tau = 20 \,[m^{-1}]$, $\mu_T = 180.32 \,[m^{-1}]$, and $\sigma_T = 15 \,[m^{-1}]$.
}\end{figure}

Figure~\ref{Fig9.1} shows the momentum spectrum averaged over different numbers of numerical runs for $\sigma_T = 15 [m^{-1}]$. For $10$ realizations (Fig.~\ref{Fig9.1}(a)), the central region ($-0.4 [m] \lesssim p_3 \lesssim 0.4 [m]$) displays sharp oscillations accompanying the largest peak, which reaches approximately $8.1 \times 10^{-13}$. These oscillations, arising from quantum interference effects also seen in individual realizations (see Fig.~\ref{Fig7}), are not fully suppressed by averaging, leaving pronounced fluctuations superimposed on an overall Gaussian-like profile. As the number of runs increases, the spectrum becomes progressively smoother and more Gaussian-like. For 50 runs [Fig.~\ref{Fig9.1}(b)], the central peak at $p_3 \approx 0$ decreases slightly to approximately $5.2 \times 10^{-13}$, while the irregular oscillations are significantly reduced. At 100 runs (Fig.~\ref{Fig9.1}(c)), the spectrum becomes smooth. The irregular fluctuations observed at smaller sample sizes are nearly suppressed. The central peak stabilizes around $5.4 \times 10^{-13}$, consistent with the 50-run case, indicating statistical convergence. Although faint ripples remain, they are nearly regular and of much smaller amplitude. At this stage, the spectrum clearly exhibits a dominant Gaussian envelope.


\par
From the above discussion of the averaged momentum spectrum, it is evident that residual oscillations persist—particularly in the central region—even after averaging over a reasonably large number of numerical runs with randomized inter-pulse delay configurations. In this sense, the spectrum is characterized by a broad Gaussian-like envelope, on top of which oscillations remain. 
To assess the convergence behavior of the averaged momentum spectrum with increasing numbers of random realizations, we quantify the statistical convergence of $\bar{f}(p_3)$ using nonlinear least-squares fits to a Gaussian model,
\begin{equation}
\bar{f}(p_3) = \frac{\mathcal{N}_0}{\sqrt{2 \pi \mathcal{S}^2}} \exp\left[ -\frac{(p_3 - \bar{p}_3)^2}{2\mathcal{S}^2} \right],
\label{eqn_guass_fit}
\end{equation}
and compute the reduced chi-squared ($\chi^2_{\text{red}}$) for each fit.
The fitted parameters and goodness-of-fit metrics are summarized in Table~\ref{tab:loggaussfits}.
\begin{table}[ht]
\centering
\caption{Fitted parameters and reduced chi-squared values for the Gaussian model applied to the averaged longitudinal momentum spectra, computed over different numbers of sample runs with randomized inter-pulse delays. The parameters $\mathcal{N}_0$, $\bar{p}_3$, and $\mathcal{S}$ represent the fitted peak amplitude, peak position, and spectral width, respectively. Quoted uncertainties correspond to 1$\mathcal{S}$ standard errors.}
\label{tab:loggaussfits}
\begin{tabular}{@{}lcccc@{}}
\toprule
Number of runs
& $\mathcal{N}_0\ (\times 10^{-13})$ 
& $\bar{p}_3(\times 10^{-4})$ 
& $\mathcal{S}$ 
& $\chi^2_{\text{red}} (\times 10^{-2})$ \\
\midrule
10 runs  & $3.8516 \pm 0.0069$ & $4.2761 \pm 1.41$   & $0.28128 \pm 0.00081$ & $8.78 $ \\
30 runs  & $3.8995 \pm 0.0039$ & $-0.7363 \pm 0.79$  & $0.28135 \pm 0.00046$ & $2.78 $ \\
50 runs  & $3.9168 \pm 0.0034$ & $1.0714 \pm 0.68$   & $0.28111 \pm 0.00039$ & $2.06$ \\
70 runs  & $3.8959 \pm 0.0028$ & $-1.1407 \pm 0.56$  & $0.28157 \pm 0.00033$ & $1.40 $ \\
100 runs & $3.8934 \pm 0.0025$ & $-1.7805 \pm 0.50$  & $0.28185 \pm 0.00029$ & $1.12 $ \\
\bottomrule
\end{tabular}
\end{table}
%
The results in Figure~\ref{Fig9.1} show the averaged momentum spectra with Gaussian fits (black dashed curves), illustrating the emergence of statistical convergence as the number of runs used for averaging. For $10$ runs, strong fluctuations remain on top of the Gaussian envelope, resulting in large peak-position uncertainty ($\bar{p}_3 = 4.2761 \pm 1.41 \times 10^{-4}$) and relatively high reduced chi-squared ($\chi^2_{red} = 8.78 \times 10^{-2}$), as summarized in Table~\ref{tab:loggaussfits}.
With 30–50 runs, the magnitude of fluctuations decreases significantly, leading to smaller parameter uncertainties and improved fit quality ($\chi^2_{red} \sim 2 \times 10^{-2}$). At 50 runs, the spectrum is smoother, the peak position approaches zero ($\bar{p}_3 = 1.0714 \pm 0.68 \times 10^{-4}$), and the width is essentially converged ($\mathcal{S} = 0.28111 \pm 0.00039$), as reported in Table~\ref{tab:loggaussfits}. For 70–100 runs, the fitted parameters show only minor changes, and the reduced chi-squared decreases further to $\sim 1 \times 10^{-2}$, demonstrating excellent agreement with the Gaussian model. At 100 runs, the spectrum becomes nearly smooth, confirming quantitative convergence and statistical robustness ($\mathcal{N}_0 = 3.893 \times 10^{-13}$, $\bar{p}_3 = -1.7805 \pm 0.50 \times 10^{-4}$, $\mathcal{S} = 0.28185 \pm 0.00029$; see Table~\ref{tab:loggaussfits}).
\par
The central Region of Interest (ROI, $-0.3 < p_3 < 0.3$) is used to quantify residual oscillations. The maximum amplitude and $\chi^2_{\rm red}$ of the Gaussian fit decrease systematically with an increasing number of runs, as summarized in Table~\ref{tab:ROI_fluctuations}: from $4.02\times 10^{-13}$ and $4.99\times 10^{-24}$ at 10 runs, to $\sim 1.2$–$2.0 \times 10^{-13}$ and $10^{-25}$–$10^{-24}$ at 70–100 runs, with the reduced chi-squared in the ROI approaching $10^{-27}$.

\begin{table}[ht]
\centering
\caption{Maximum amplitude of oscillation and reduced chi-squared values in the central ROI ($-0.3 < p_3 < 0.3$) for different numbers of runs.}
\label{tab:ROI_fluctuations}
\begin{tabular}{@{}lcc@{}}
\toprule
Number of runs & Max amplitude ($\times 10^{-13}$) & $\chi^2_{\rm red}$ ($\times 10^{-26}$) \\
\midrule
10  & 4.0248 & 1.6809 \\
30  & 1.7816 & 0.4525 \\
50  & 1.5648 & 0.3568\\
70  & 2.0040 & 0.3105\\
100 & 1.1815 & 0.2053\\
\bottomrule
\end{tabular}
\end{table}
Therefore, the results clearly demonstrate that increasing the number of runs improves statistical convergence. The spectral fluctuations are progressively suppressed, parameter uncertainties shrink, and the Gaussian model provides an increasingly reliable description of the underlying momentum distribution.  By 100 runs, the momentum distribution is statistically robust, and the Gaussian model accurately represents the underlying spectrum.


\begin{figure}[tbp]
\suppressfloats
\centering
{\includegraphics[width=0.968\columnwidth]{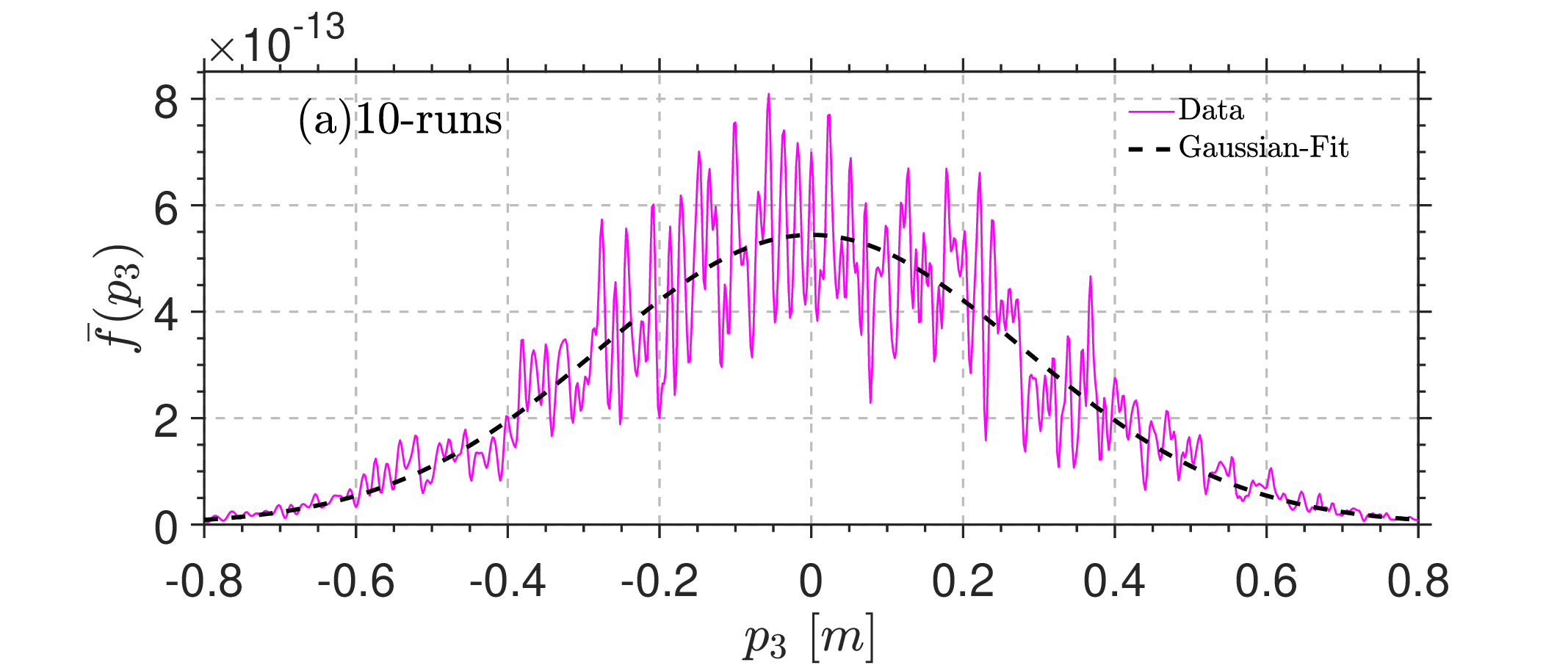}
\includegraphics[width=0.968\columnwidth]{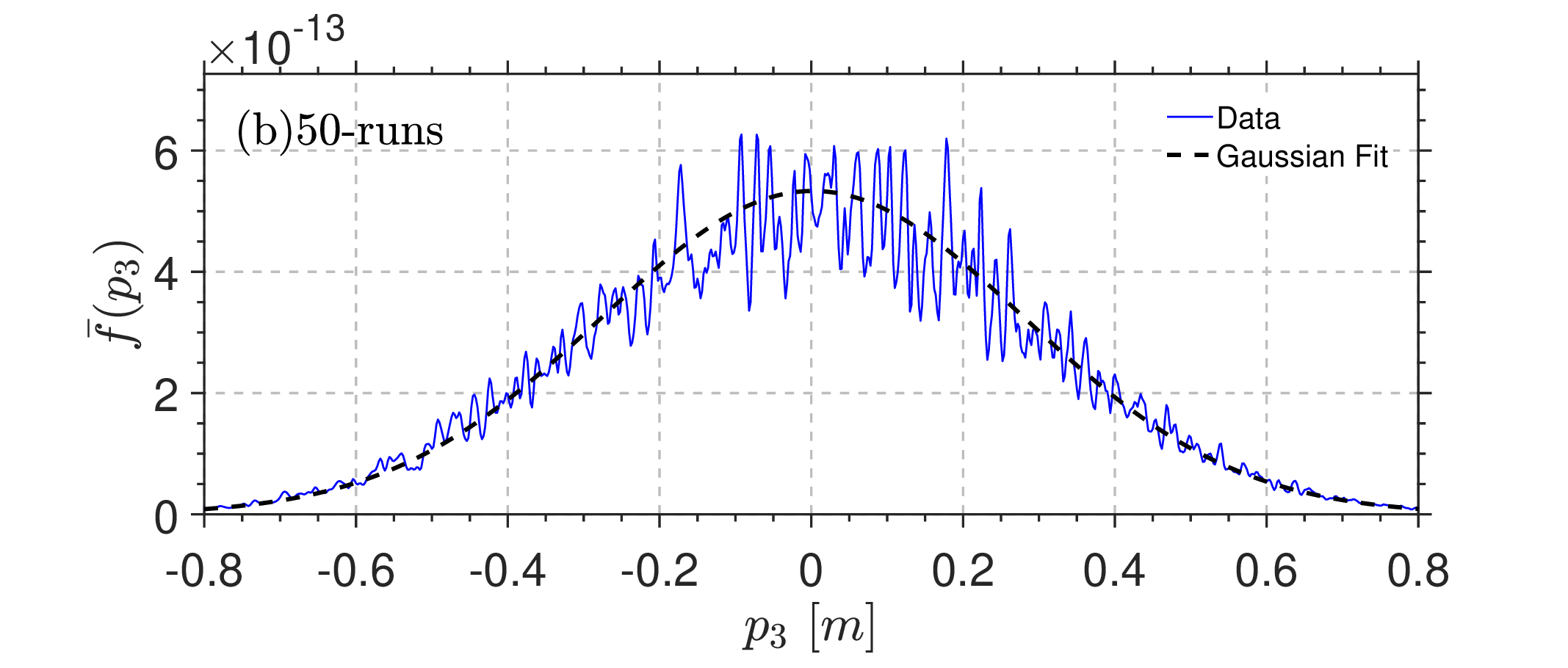}
\includegraphics[width=0.968\columnwidth]{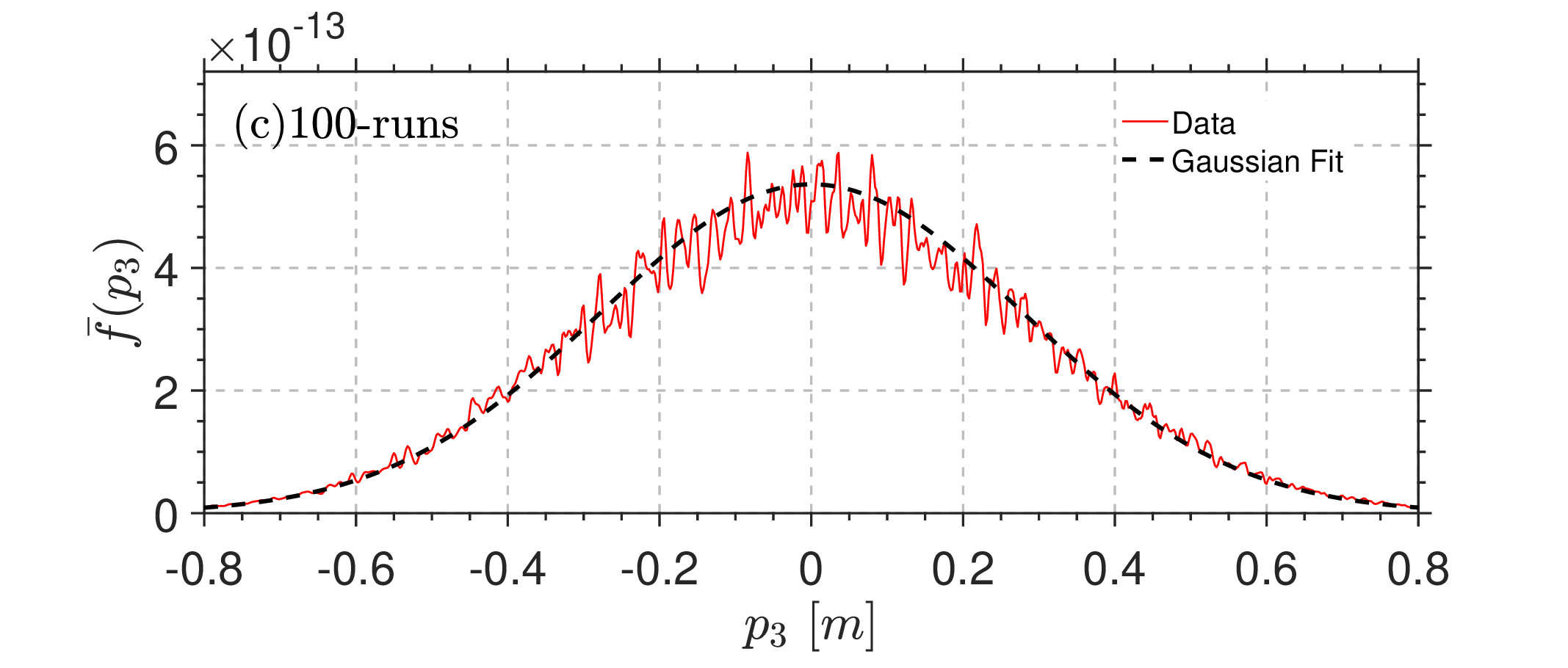}}
\caption{\label{Fig9.2}
 Same as in Fig.~\ref{Fig9.1}, except for an alternating-sign four-pulse electric field with  $\sigma_T = 45 \,[m^{-1}]$.}
 \end{figure}


Next, we consider the case with a stronger degree of randomness, $\sigma_T = 45[m^{-1}]$. Figure~\ref{Fig9.2} displays the averaged longitudinal momentum spectra $\bar{f}(p_3)$ for this value of $\sigma_T$, corresponding to stronger temporal disorder.
 Similar to the $\sigma_T = 15[m^{-1}]$ case (Fig.~\ref{Fig9.1}), strong oscillatory behaviour persists when averaging over only a few realizations (10 runs). These oscillations are gradually suppressed as the ensemble size increases, and the distribution converges toward a Gaussian-like profile with only small residual oscillations when 100 runs are included. 
 Representative results for $10$, $50$, and $100$ runs are shown in the three panels of Fig.~\ref{Fig9.2}. In each case, the spectra are fitted with the Gaussian model in Eq.~\eqref{eqn_guass_fit}, which captures the overall envelope of the distributions. The extracted fit parameters, along with the reduced chi-squared values, are listed in Table~\ref{tab:loggaussfits2}, providing a quantitative measure of convergence. 
 For 10 runs [Fig.~\ref{Fig9.2}(a)], the spectrum remains highly irregular, with strong fluctuations about the Gaussian envelope. From Table~\ref{tab:loggaussfits2}, the fitted parameters are $\mathcal{N}_0 = 3.8137 \times 10^{-13}$, 
$\bar{p}_3 = 2.59 \times 10^{-4}$, and $\mathcal{S} = 0.27947$, while the reduced chi-squared, $\chi^2_{\text{red}} = 8.99 \times 10^{-2}$, indicates that the Gaussian model captures only the broad trend but not the fine details. Thus, averaging over 10 runs is insufficient for statistical reliability.
At 50 runs [Fig.~\ref{Fig9.2}(b)], the oscillation amplitudes are much reduced, and the spectrum closely resembles a Gaussian. The parameters stabilize to $\mathcal{N}_0 = 3.733 \times 10^{-13}$, $\bar{p}_3 = 2.126 \times 10^{-3}$, and $\mathcal{S} = 0.27924$, with a reduced chi-squared of $2.25 \times 10^{-2}$, signaling a marked improvement in the fit quality.
Finally, for 100 runs [Fig.~\ref{Fig9.2}(c)], the averaged spectrum becomes very smooth, with residual oscillations strongly suppressed. The Gaussian model provides an excellent fit, with nearly saturated parameters $\mathcal{N}_0 = 3.7629 \times 10^{-13}$, $\bar{p}_3 = 5.69 \times 10^{-4}$, and $\mathcal{S} = 0.27989$, while the reduced chi-squared drops to $1.05 \times 10^{-2}$. This confirms that statistical convergence is achieved at large ensemble sizes.

\begin{table}[h!]
\centering
\caption{Fitted parameters and reduced chi-squared values for the Gaussian model applied to the averaged longitudinal momentum spectra, computed over different numbers of ensemble runs with randomized time delays. Quoted uncertainties correspond to 1$\mathcal{S}$ standard errors.}
\label{tab:loggaussfits2}
\begin{tabular}{@{}lcccc@{}}
\toprule
Number of runs
& $\mathcal{N}_0\ (\times 10^{-13})$ 
& $\bar{p}_3(\times 10^{-4})$ 
& $\mathcal{S}$ 
& $\chi^2_{\text{red}}(\times 10^{-2})$ \\
\midrule
10 runs  & $3.8137 \pm 0.0065$ & $2.5916 \pm 1.32$   & $0.27947 \pm 0.00073$ & $8.99$ \\
30 runs  & $3.8644 \pm 0.0037$ & $9.9292 \pm 0.74$   & $0.27881 \pm 0.00041$ & $2.88$ \\
50 runs  & $3.7332 \pm 0.0032$ & $21.257 \pm 0.66$   & $0.27924 \pm 0.00036$ & $2.25$ \\
70 runs  & $3.8164 \pm 0.0027$ & $7.9331 \pm 0.55$   & $0.28018 \pm 0.00030$ & $1.55$\\
100 runs & $3.7629 \pm 0.0022$ & $5.6917 \pm 0.45$   & $0.27989 \pm 0.00025$ & $1.05$ \\
\bottomrule
\end{tabular}
\end{table}


Residual oscillations in the central ROI are quantified in Table~\ref{tab:ROI_fluctuations2}. At 10 runs, the largest fluctuation reaches $2.95 \times 10^{-13}$ with $\chi^2_{ROI} = 4.02 \times 10^{-24}$. Increasing the ensemble size steadily suppresses these deviations: the fluctuation amplitude drops below $2.0 \times 10^{-13}$ for 30–50 runs and reaches $1.15 \times 10^{-13}$ at 100 runs, with $\chi^2_{\rm ROI}$ correspondingly reduced to $5.36 \times 10^{-25}$ ($\chi^2_{\rm red} = 1.81 \times 10^{-27}$). This monotonic improvement confirms that ensemble averaging systematically damps residual oscillations and yields statistically converged Gaussian spectra, though weak fluctuations remain visible even at 100 runs.

\begin{table}[ht]
\centering
\caption{Residual oscillations in the central ROI ($-0.3 < p_3 < 0.3$) for $\sigma_T = 45~[m^{-1}]$. Listed are the maximum amplitude of oscillation and the reduced chi-squared $\chi^2_{\rm red}$ for different ensemble sizes.}
\label{tab:ROI_fluctuations2}
\begin{tabular}{@{}lcc@{}}
\toprule
Runs & Max amplitude ($\times 10^{-13}$) & $\chi^2_{\rm red}$ ($\times 10^{-26}$) \\
\midrule
10   & 2.9483 & 1.3519 \\
30   & 1.8882 & 0.4341 \\
50   & 1.8226 & 0.4677 \\
70   & 1.4549 & 0.3129 \\
100  & 1.1525 & 0.1805 \\
\bottomrule
\end{tabular}
\end{table}

\par
The averaged momentum spectra can then be directly compared with the non-stochastic case ($\sigma_T = 0$). In this deterministic limit, the momentum distribution displays sharply resolved interference fringes, a clear manifestation of quantum coherence in the time domain (see Fig.~\ref{Fig0}. By contrast, introducing temporal randomness in the pulse sequence ($\sigma_T > 0$) progressively degrades this coherence. For low disorder ($\sigma_T = 15[m^{-1}]$), ensemble averaging suppresses fine oscillatory structures, producing a smoother spectral profile. At high disorder ($\sigma_T = 45[m^{-1}]$), coherence is largely destroyed, and the spectrum evolves into a broad Gaussian-like envelope with a central peak at $p_3 \sim 0$ and exponential-like decay in the tails.
Small residual oscillations remain in the central region, indicating that partial coherence persists even under significant randomness. These comparisons clearly show that the strength of timing disorder effectively governs the crossover from coherent, interference-dominated spectra to incoherent, Gaussian-like distributions. Ensemble averaging becomes especially important in realistic experimental settings where laser pulses exhibit intrinsic jitter. Averaging over approximately $50-100$  realizations is therefore essential for extracting physically meaningful and reproducible features in the presence of shot-to-shot variations~\cite{Macias:2021ume}. 



\begin{figure}[tbp]\suppressfloats
	\centering
	{
		\includegraphics[width=0.99\columnwidth]{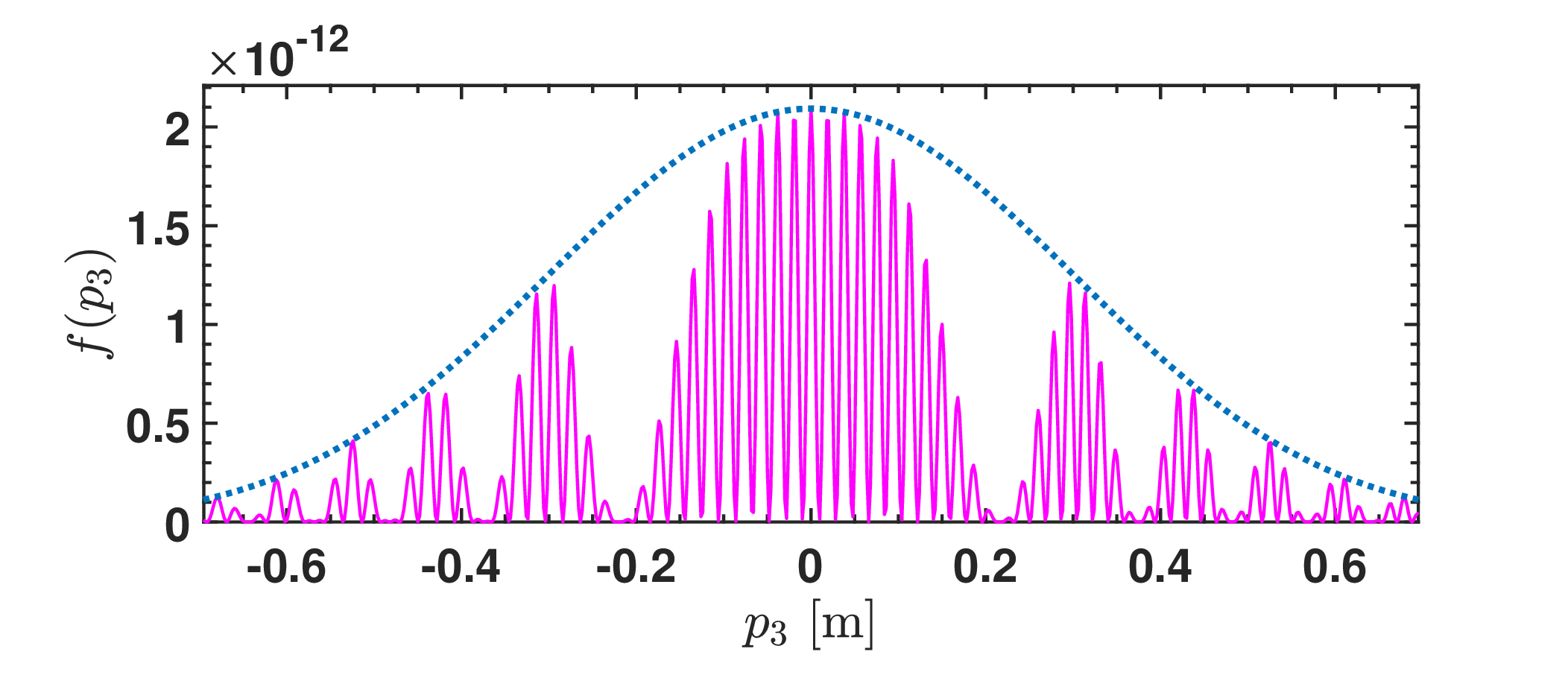}
		\includegraphics[width=0.99\columnwidth]{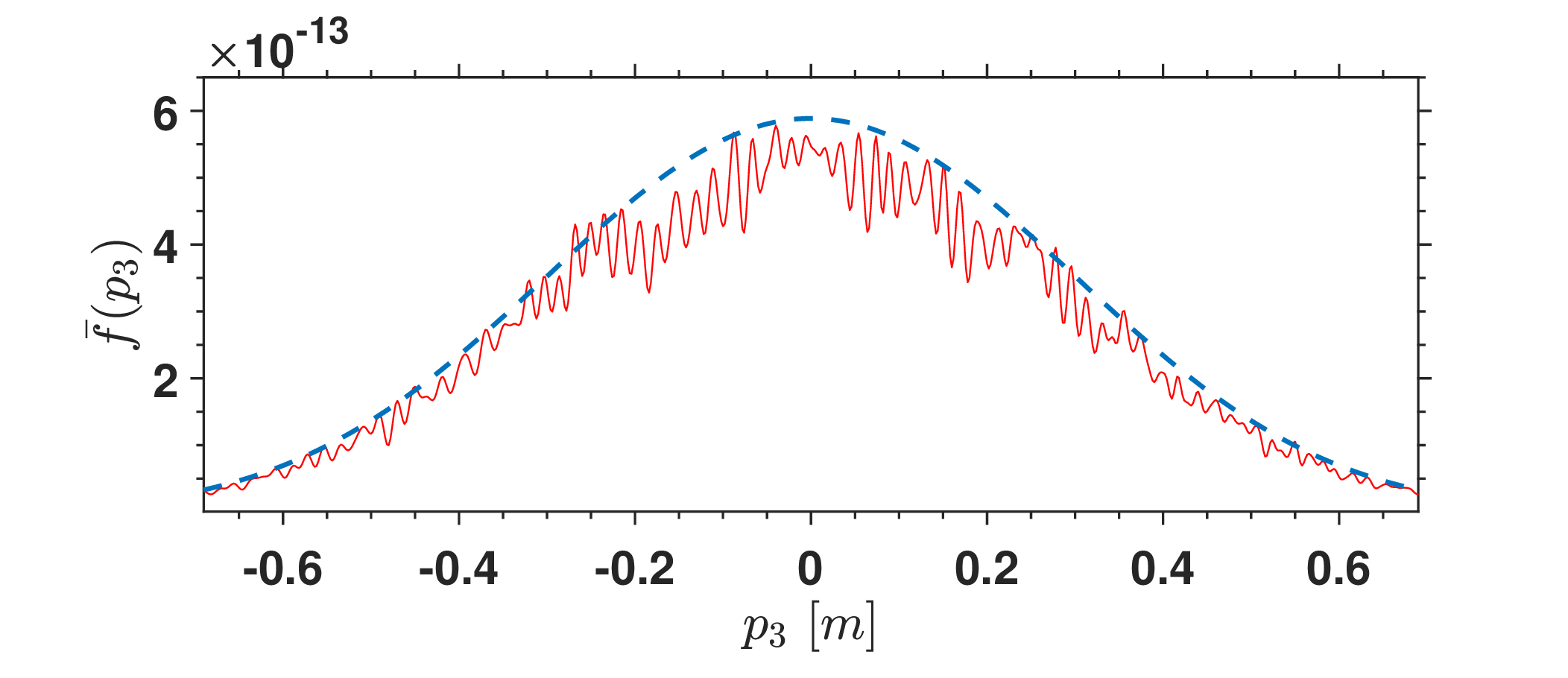}
		\includegraphics[width=0.99\columnwidth]{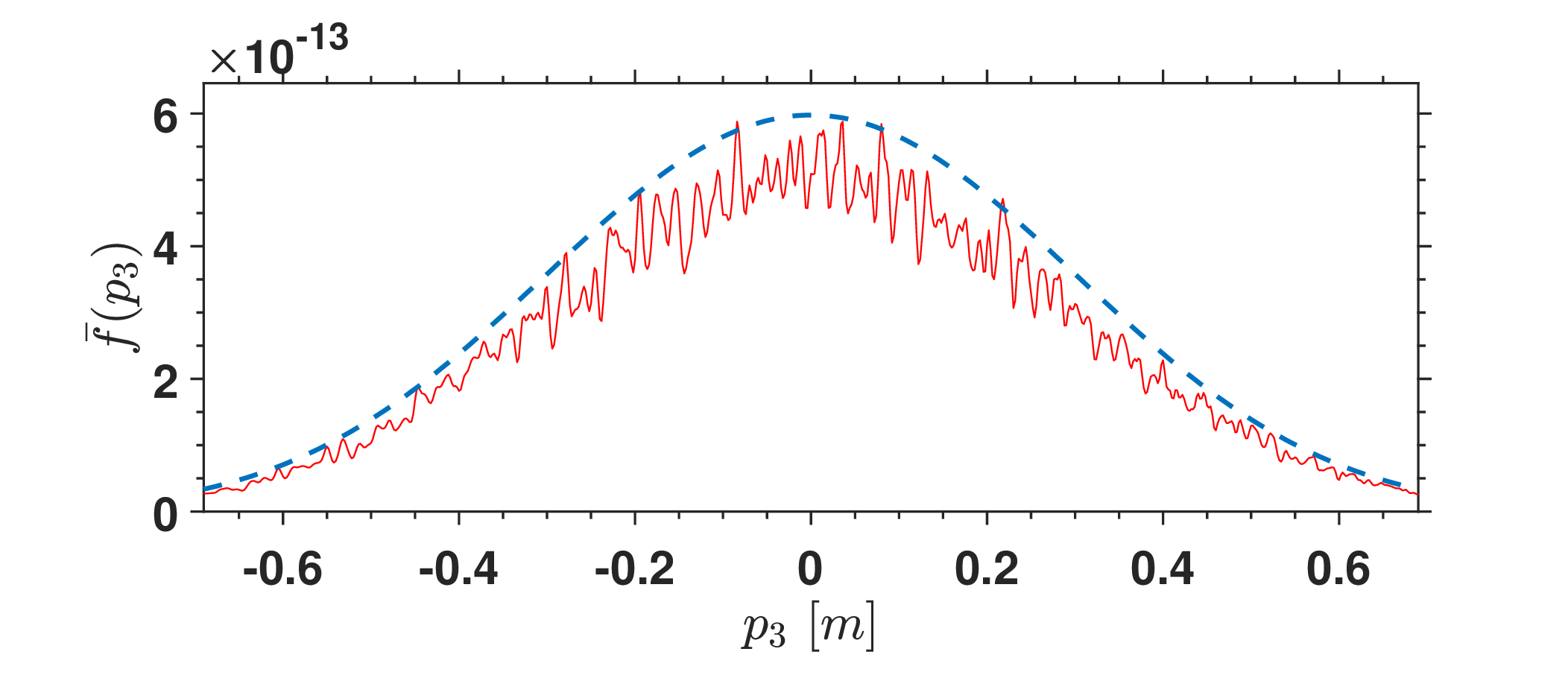}}
	\caption{
		\label{benchmark}\textbf{Upper panel:} Longitudinal momentum spectrum for the non-stochastic case ($\sigma_T = 0$). The dashed blue curve is $N^2 = 16$ times the single-pulse spectrum, illustrating the coherent $N^2$ enhancement. 
		\textbf{Middle panel:} Ensemble-averaged spectrum over 100 realizations for $\sigma_T = 15 [m^{-1}]$. The dashed blue curve is $(2N+1)/2 \approx 4.5$ times the single-pulse result, indicating partial loss of coherence. 
		\textbf{Lower panel:} Ensemble-averaged spectrum over 100 realizations for $\sigma_T = 45[m^{-1}]$. The dashed blue curve again follows $(2N+1)/2 \approx 4.5$ times the single-pulse spectrum, confirming the transition to an incoherent sum of pulse contributions.
	}
\end{figure}

 
\begin{figure}[tbp]\suppressfloats
\centering
{
\includegraphics[width=0.954\columnwidth]{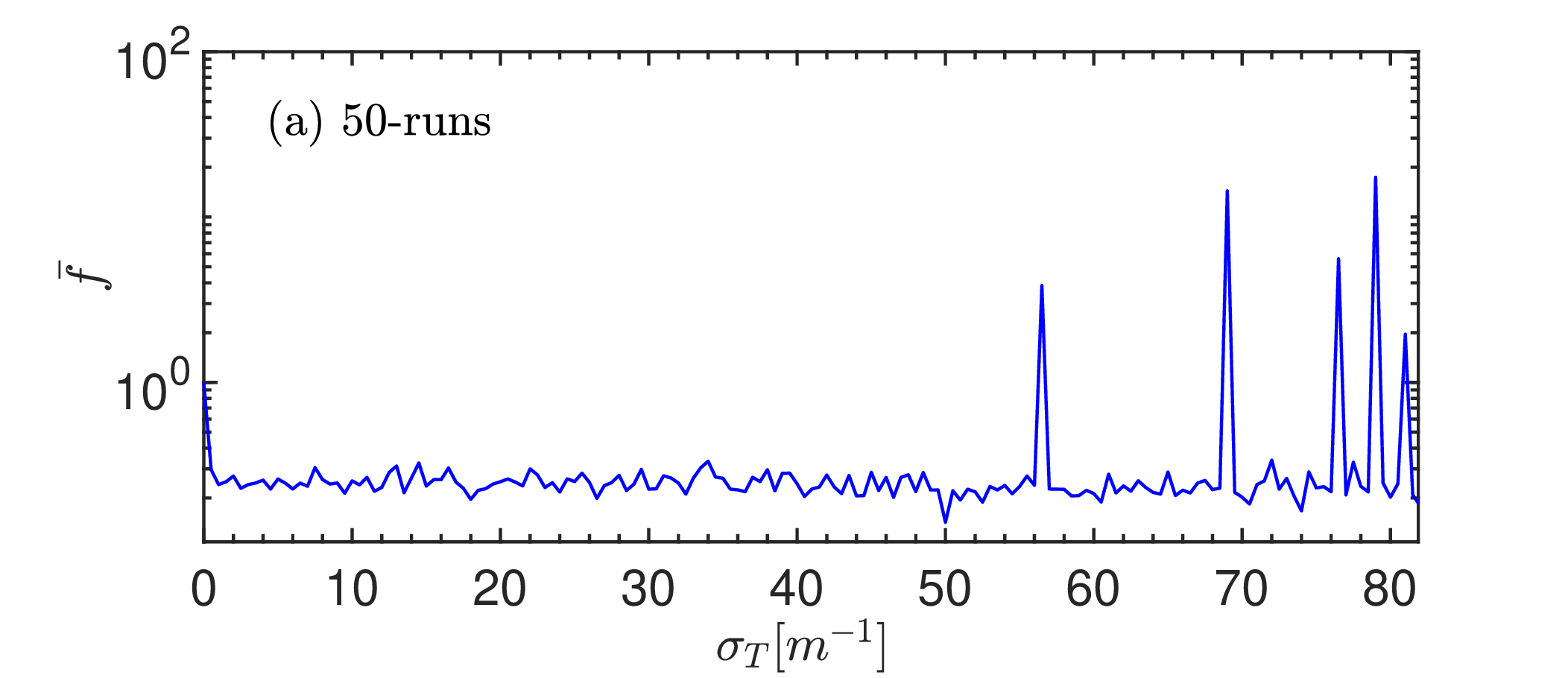}
\includegraphics[width=0.954\columnwidth]{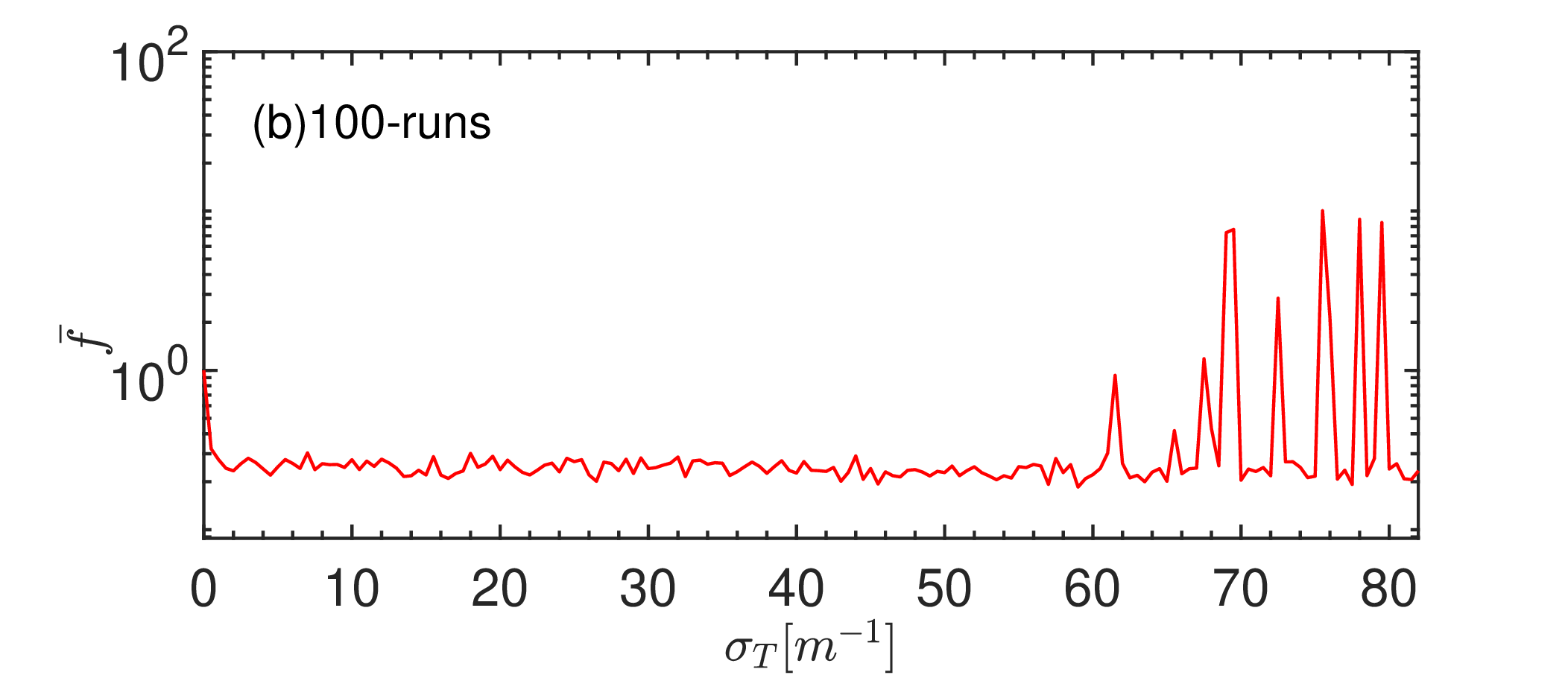}}
\caption{\label{f_mean_4}
Averaged distribution function $\bar{f}$ at zero momentum, computed over multiple numerical runs with randomized inter-pulse delays, as a function of the randomness parameter $\sigma_T$ for an alternating-sign $N$-pulse electric field $E(t)$ with $N=4$. The values are normalized to the corresponding result for the non-stochastic case ($\sigma_T = 0$). The field parameters are $E_0 = 0.1E_c$, $\tau = 20,[m^{-1}]$, and $\mu_T = 180.32,[m^{-1}]$.}
\end{figure}




Figure~\ref{benchmark} serves as a predictive benchmark that illustrates the transition from coherent to incoherent pair production as timing randomness increases. In the absence of randomness, the spectrum exhibits a sharp $N$-slit interference pattern, with the central peak enhanced by a factor of $N^2$ relative to a single Sauter pulse—a hallmark of fully constructive quantum interference~\cite{Akkermans:2011yn}. This scaling is evident in the upper panel, where the dashed blue curve matches $N^2 = 16$ times the single-pulse momentum distribution.

Introducing random delays randomizes the relative quantum phases between successive pulses. For individual stochastic realizations, the interference pattern becomes distorted and asymmetric (Figs.~\ref{Fig7}--\ref{Fig9}). When averaged over many such realizations, the interference fringes—which occur at different momenta in each run—average out, leaving an incoherent sum of contributions from the $N$ individual pulses. This process is analogous to the central limit theorem, leading to a Gaussian-like envelope in the averaged spectrum.

The fitted Gaussian parameters (Tables~\ref{tab:loggaussfits} and \ref{tab:loggaussfits2}) support this interpretation: the spectral width $\mathcal{S}$ is consistent with that of a single Sauter pulse of duration $\tau = 20\,[m^{-1}]$ and amplitude $E_0 = 0.1E_c$. Thus, in the high-disorder limit, the averaged spectrum converges to approximately $N$ times the single-pulse momentum distribution.

In the middle and lower panels of Fig.~\ref{benchmark}, the dashed blue curve scales as $(2N+1)/2 \approx 4.5$ times the single-pulse result (for $N=4$), rather than $N^2$. This reduced scaling factor—close to $N$ rather than $N^2$—quantitatively demonstrates the destruction of phase coherence by random time delays. The transition from $N^2$ to $\sim N$ scaling reflects the shift from constructive interference of amplitudes to incoherent addition of probabilities.
The emergence of a Gaussian-like envelope in the ensemble-averaged spectra is a direct signature of decoherence induced by timing disorder. It provides a clear link between the stochastic multi-pulse field and the underlying single-pulse momentum distribution, confirming that in the highly stochastic regime, pair production reduces to an incoherent sum of independent pulse contributions. Akkermans and Dunne ~\cite{Akkermans:2011yn} demonstrated that in a regular alternating-sign pulse train, the central peak scales as $N^2$, making such configurations promising for enhancing Schwinger pair production. Our results extend this picture by showing that when randomness is introduced, the scaling transitions from $N^2$ to approximately $N$, corresponding to the loss of quantum coherence. This insight is crucial for designing future experiments where timing jitter is unavoidable.


\begin{figure}[tbp]\suppressfloats
\centering
{
\includegraphics[width=0.99\columnwidth]{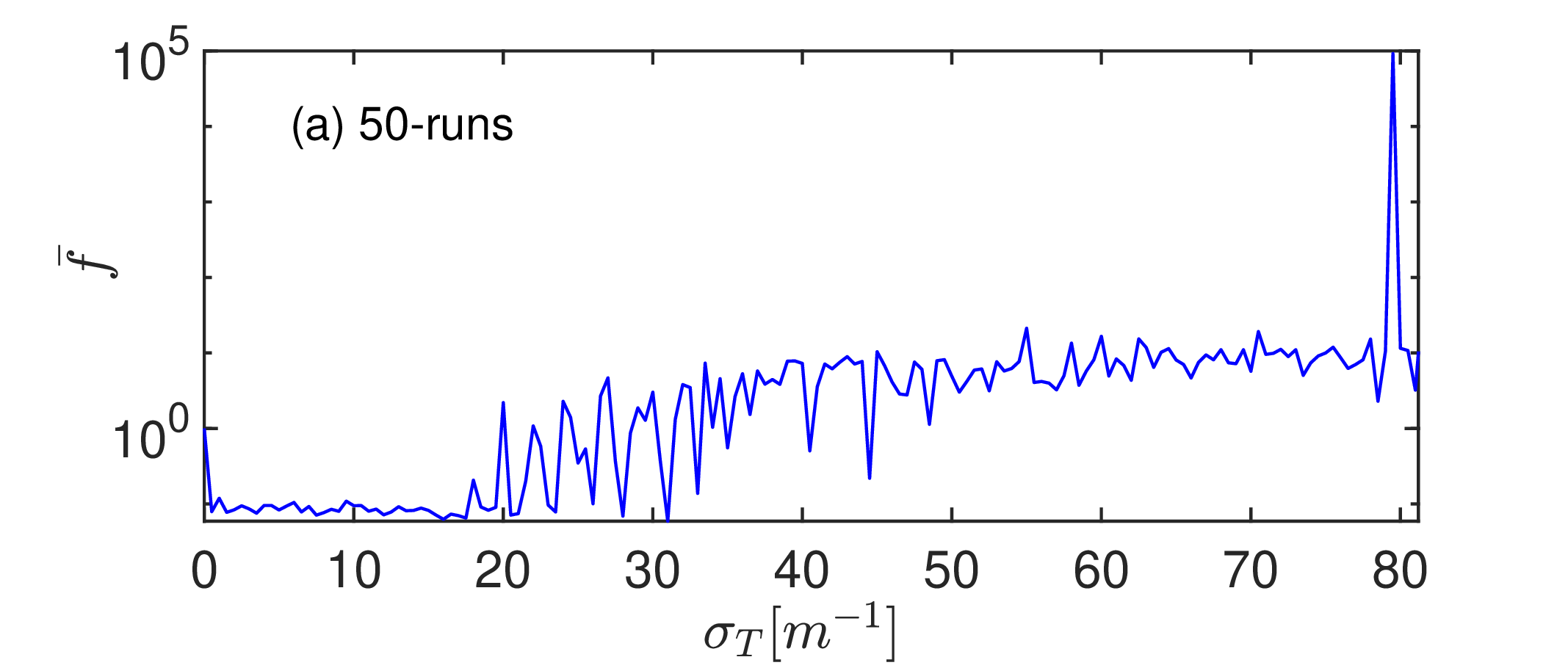}
\includegraphics[width=0.99\columnwidth]{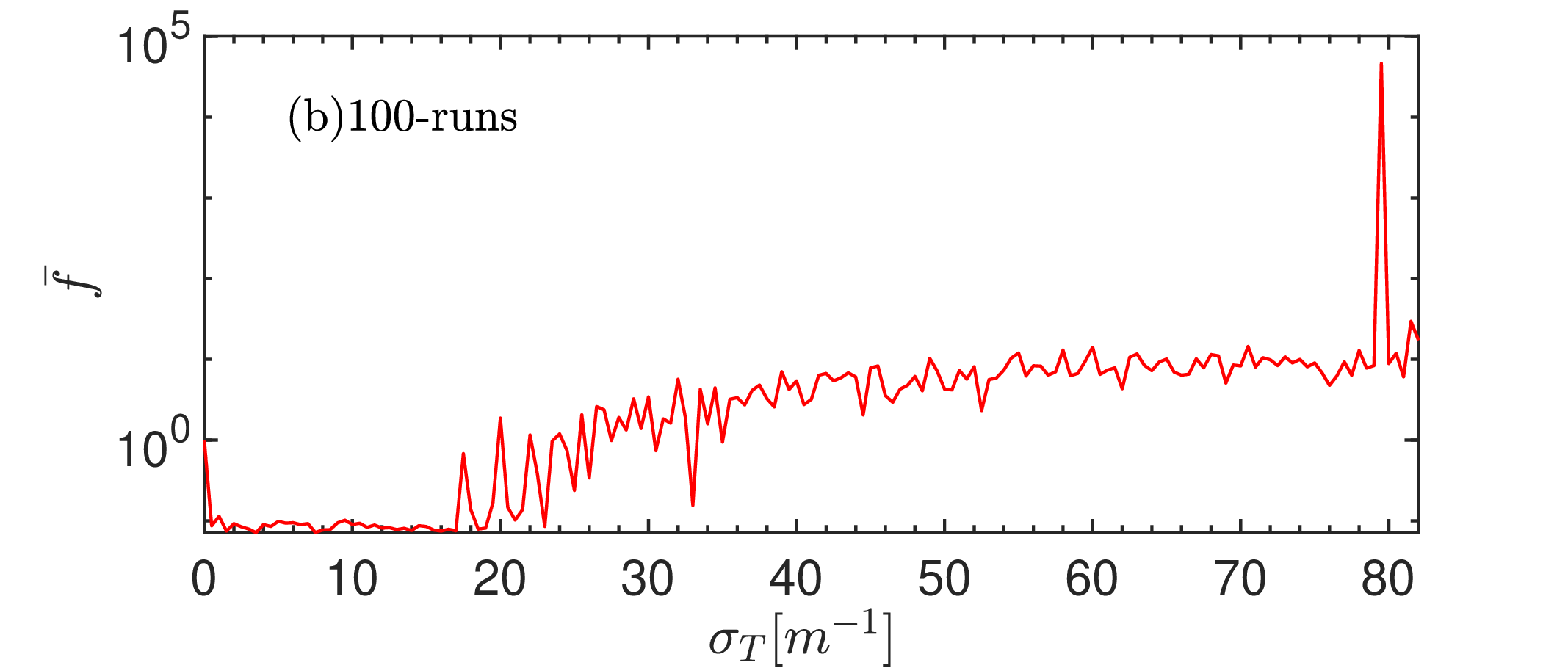}}
\caption{\label{f_mean_12}The same as in Fig.~\ref{f_mean_4}, except for an alternating-sign $N$-pulse electric field with $N=12.$}
\end{figure}




\par
We now extend our analysis to examine the role of randomness in shaping the momentum spectrum of created EPPs. Increasing stochasticity in the inter-pulse delays generally degrades the interference fringes; however, the central peak of the spectrum remains especially sensitive to such variations. To study this effect, we focus on the distribution function at $\bm{p}=0$, which corresponds to the central peak in the non-stochastic limit and has been highlighted in earlier studies \cite{Akkermans:2011yn} as the point where the $N^2$ enhancement in pair production is concentrated. Motivated by this, we analyze how the distribution at $p=0$ evolves with the degree of randomness, parameterized by the standard deviation $\sigma_T$. For this purpose, we consider the averaged distribution function at zero momentum, $\bar{f}(\bm{p}=0)$, as a function of $\sigma_T$, computed over ensembles of 50 and 100 realizations. Such averaging is crucial in experimental situations involving stochastic pulse sources, where multiple laser shots are accumulated, and it is the averaged behavior that corresponds to the measured signal. 

\par
Figure~\ref{f_mean_4} shows $\bar{f}(\bm{p}=0)$, normalized to its non-stochastic value ($\sigma_T=0$), for a four-pulse sequence. Panels (a) and (b) present the results for averages over 50 and 100 configurations, respectively. At small $\sigma_T$, the averaged distribution function remains suppressed and nearly constant. As $\sigma_T$ increases, irregular variations develop, and around $\sigma_T \approx 70\,[m^{-1}]$ sharp peaks appear in both panels. These peaks indicate that certain levels of timing randomness can lead to enhanced distribution function at zero momentum,$\bar{f}$. Their reproducibility with larger ensembles confirms that they are robust physical features rather than statistical fluctuations. Overall, increasing randomness tends to suppress the central peak, yet for specific values of $\sigma_T$, the averaged distribution at $p=0$ can still be enhanced. This trend is consistent with the residual oscillatory behavior observed in the momentum spectrum near $p_3 =0$, which survives averaging over many random realizations.
\par

To explore how the number of pulses affects the sensitivity to timing randomness, we now consider larger pulse trains with $N=12$ and $N=20$.
Figure~\ref{f_mean_12} shows the averaged distribution function $\bar{f}(\bm{p}=0)$, normalized to its value at $\sigma_T = 0$, for the case of $N=12$ pulses. The results are averaged over (a) 50 and (b) 100 random configurations of the inter-pulse delays. 
For small $\sigma_T$, the distribution is strongly suppressed compared to the non-stochastic case, reflecting more pronounced destructive interference than for $N=4$ scenario (Fig.~\ref{f_mean_4}). As $\sigma_T$ increases beyond $20[m^{-1}]$, fluctuations in $\bar{f}$ begin to appear, and sharp peaks are visible across a broad range of $\sigma_T$. In contrast to the $N=4$ case, where only a few peaks emerged at large $\sigma_T$, the $N=12$ case exhibits a much richer structure, with many peaks distributed throughout $20 \lesssim \sigma_T \lesssim 70[m^{-1}]$. In this interval, the averaged values fluctuate around an order of magnitude above the non-stochastic baseline.
At larger randomness, particularly near $\sigma_T \approx 80[m^{-1}]$, a sharp enhancement appears. With 50 runs, this feature is already visible, but with 100 runs (panel b), it becomes clear and reproducible, showing an increase of more than four orders of magnitude. This peak originates from rare but favorable time-delay configurations and only emerges distinctly when averaging over sufficiently large ensembles. Overall, Fig.~\ref{f_mean_12} shows that for $N=12$ the system is significantly more sensitive to stochastic delays than for smaller pulse numbers. 
\begin{figure}[tbp]\suppressfloats
\centering

\includegraphics[width=0.954\columnwidth]{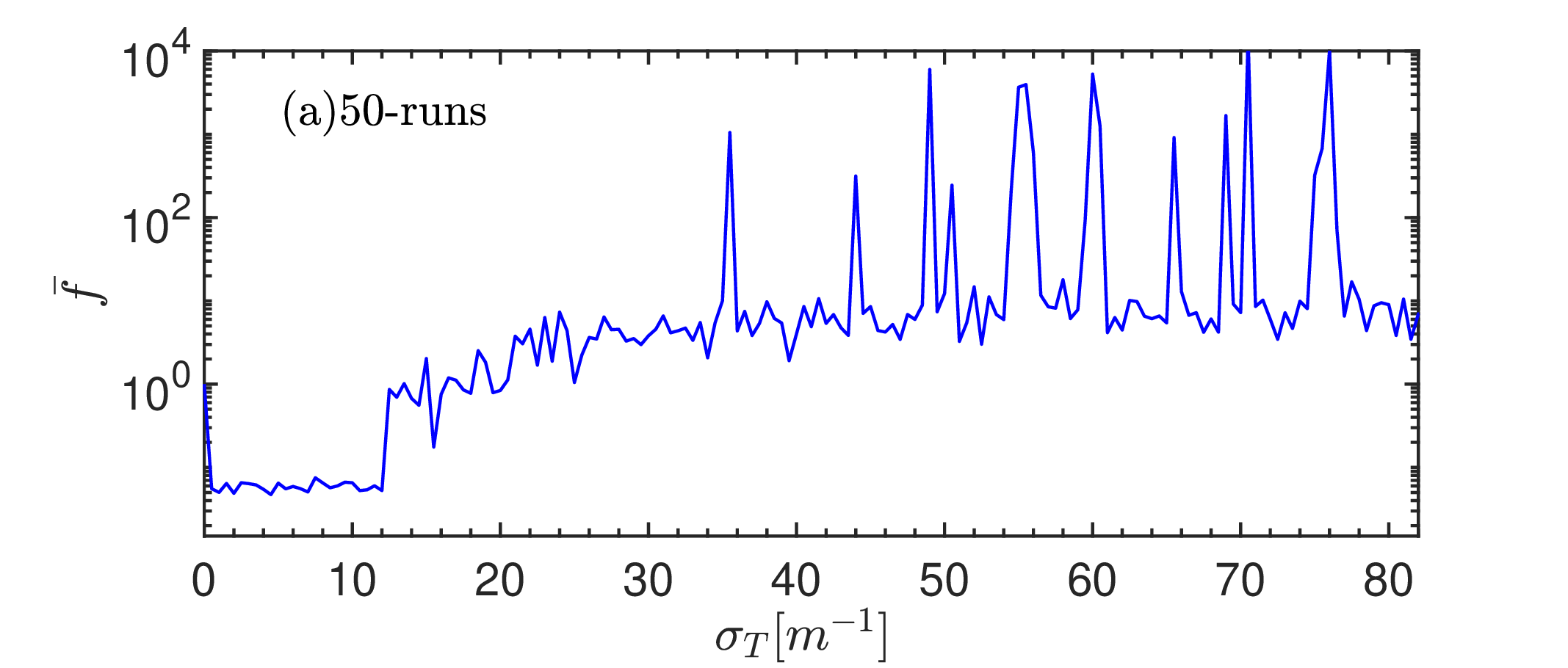}
\includegraphics[width=0.954\columnwidth]{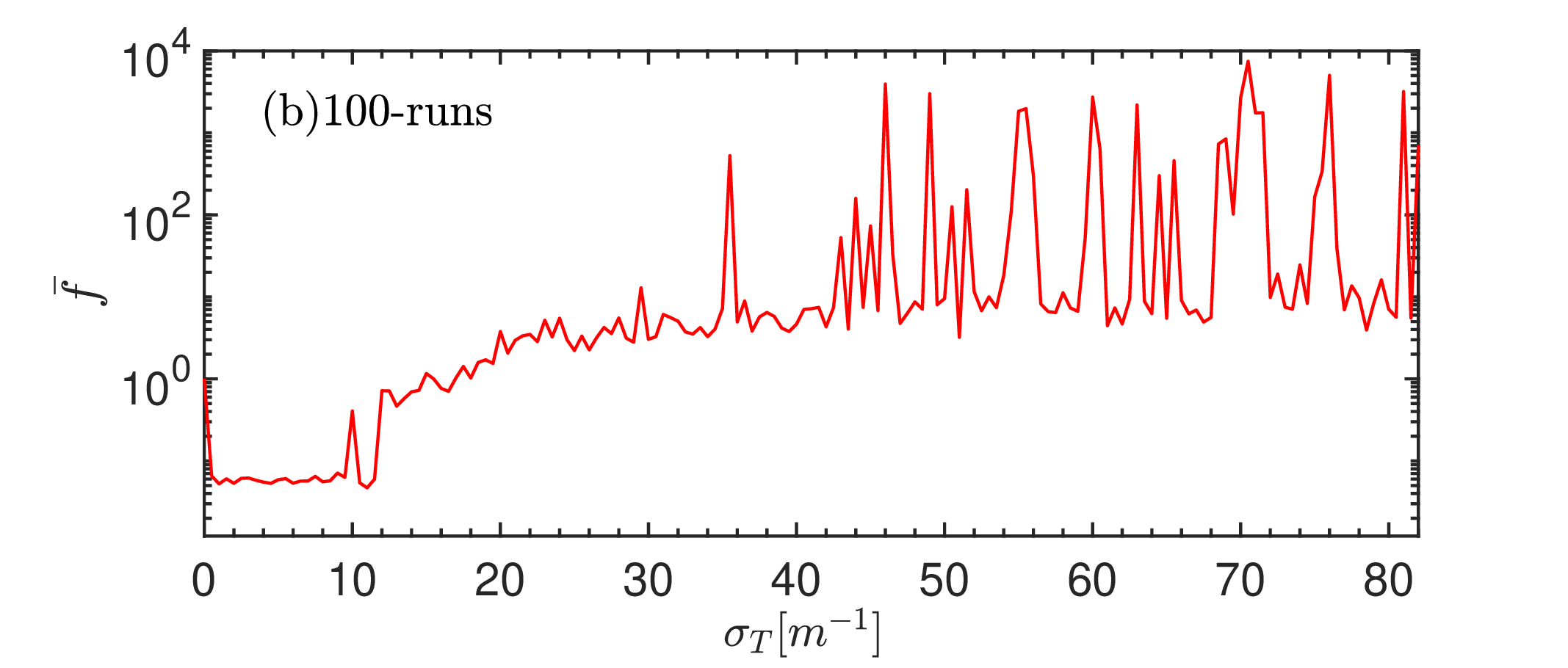}
\caption{\label{f_mean_20}The same as in Fig.~\ref{f_mean_4}, except for an alternating-sign $N$-pulse electric field with $N=20$.}
\end{figure}
\par
We now consider a multi-pulse train with $N=20$ pulses, doubling the number used in the previous case ($N=12$). Figure~\ref{f_mean_20} shows the averaged distribution function $\bar{f}(p=0)$, normalized to its value at $\sigma_T=0$, as a function of the time-delay randomness parameter $\sigma_T$. For small $\sigma_T$, the nearly regular delays produce structured oscillations in $\bar{f}$. As $\sigma_T$ increases, the timing becomes irregular, and the distribution develops noisy fluctuations that strongly depend on each random realization. With 50 runs (panel a), the distribution remains suppressed at low $\sigma_T$, then rises with growing randomness, showing distinct peaks beyond $\sigma_T \sim 40[m^{-1}]$. With 100 runs (panel b), these peaks become sharp and reproducible, confirming that they are genuine features. Across all panels, $\bar{f}$ shows a clear progression as the randomness parameter increases. For $\sigma_T \lesssim 10$, the average already exceeds the $\sigma_T=0$ value. In the intermediate range, $10 \lesssim \sigma_T \lesssim 40$, alternating maxima and minima appear, with enhancements reaching nearly tenfold. At larger values of $\sigma_T$, the amplification becomes much stronger, with the averaged distribution rising by almost three orders of magnitude around $\sigma_T \approx 50[m^{-1}]$.
Very sharp peaks also occur at certain values, arising from rare delay configurations that yield exceptionally strong particle production.
 Compared to $N=12$ (Fig.~\ref{f_mean_12}), the $N=20$ case shows 
 stronger amplification and clearer trends, while for $N=4$ only a few peaks are visible. This progression highlights how larger pulse numbers enhance the sensitivity to timing randomness.


The emergence of sharp peaks in \(\bar{f}(\bm{p} = 0)\) at large \(\sigma_T\) can be understood through statistical sampling of delay configurations. At low \(\sigma_T\) (coherent regime), the pulse train is nearly regular, and quantum phases are locked,especially for larger \(N\), where the normalized \(\bar{f}(0)\) starts well below unity. As \(\sigma_T\) increases into the intermediate regime, phase coherence is broken and contributions from pulses add incoherently, yielding a gradual rise in \(\bar{f}(0)\) without sharp features. At high \(\sigma_T\), however, the broad distribution of delays creates a vast ``search space'' of possible pulse sequences. Within this space, random sampling occasionally generates rare, optimized configurations where the relative delays coincidentally align to produce strong constructive interference—more efficient than the simple incoherent sum.

For \(N=4\), a relatively large \(\sigma_T\) (\(\approx 60\text{--}70 [m^{-1}]\)) is needed to provide a sufficiently wide parameter space for these optimal configurations to appear with statistical significance in the ensemble average. In contrast, for larger \(N\) (12, 20), the system starts in a deeper interference minimum, making it more sensitive to randomness. Moreover, with more pulses, there are more combinatorial possibilities for creating highly constructive sequences. Consequently, the threshold \(\sigma_T\) for observing sharp, order-of-magnitude enhancements is lower (\(\approx 30\text{--}50 [m^{-1}]\)), as seen in Figs.~\ref{f_mean_12} and \ref{f_mean_20}.

Thus, the peaks reflect a stochastic optimization mechanism where randomness, at specific strengths, maximizes the probability of generating pulse trains that strongly enhance pair production.
This insight suggests that tailored randomness could be strategically exploited to optimize pair production in experimental settings where perfect timing control is challenging.


The multi-pulse train with random delays explores an effectively infinite space of possible pulse timing arrangements. Our results demonstrate that
increasing randomness can occasionally ``find'' rare configurations that strongly enhance pair production. This naturally raises the question: what is the optimal temporal arrangement of pulses that maximizes pair creation?
While our study shows that such optimized configurations exist and can be accessed stochastically, a systematic search for the absolute optimum and a deeper analytical understanding of the underlying interference conditions go beyond the scope of the present work. This constitutes a compelling direction for future research.

\section{Conclusions} \label{summary}
We investigated the creation of EPPs in a sequence of alternating-sign Sauter-like pulses with randomized inter-pulse delays, modeled by a Gaussian distribution with standard deviation $\sigma_T$ controlling the degree of temporal disorder. For $N=4$, the longitudinal momentum spectra exhibit a clear progression with increasing $\sigma_T$. In the deterministic limit ($\sigma_T=0$), a regular $N$-slit interference pattern emerges, characterized by a dominant central band and symmetric side fringes with high visibility.
For low randomness ($\sigma_T=15[m^{-1}]$), the central broad band fragments into sub-bands, distorting the structure into irregular, asymmetric peaks and inducing a left–right asymmetry across runs. At moderate disorder ($\sigma_T=45[m^{-1}]$), the continuous fringe pattern dissolves into clusters of irregular peaks, accompanied by suppressed side bands and enhanced run-to-run fluctuations. For strong randomness ($\sigma_T=75[m^{-1}]$), the fringe-like interference pattern becomes almost completely disordered: the central region is densely populated with erratic peaks, and the notion of a band-like structure disappears entirely. Stochastic fluctuations, with pronounced run-to-run variability, dominate the resulting distribution. Taken together, these results demonstrate that increasing temporal randomness progressively degrades fringe-like patterns arising from quantum interference. While residual constructive interference persists at intermediate values of $\sigma_T$, the spectrum ultimately becomes dominated by irregular fluctuations as randomness grows, signaling a transition from a coherent interference-dominated regime to one governed by stochastic behavior. To obtain statistically reliable results, we performed ensemble averaging over multiple realizations, particularly for $N=4$. The averaged momentum spectra exhibit a broad Gaussian-like envelope with residual oscillatory features, while central-region fluctuations persist even after averaging. These findings are especially relevant for realistic experimental conditions, where averaging over multiple laser shots is necessary.
\par
Interestingly, beyond the general trend of fringe-like interference pattern modification and suppression, we also observe that randomness can induce a pronounced enhancement of the central peak in the momentum spectrum. Specifically, the distribution function at zero momentum shows a fluctuating dependence on $\sigma_T$, with significant amplification at higher disorder. For $N=12$, a noticeable enhancement appears around $\sigma_T \sim 40[m^{-1}]$, whereas for smaller pulse numbers($N=4$), a nearly tenfold increase is observed only at larger disorder, $\sigma_T \sim 70[m^{-1}]$. This progression with increasing $N$ highlights how larger pulse trains enhance the sensitivity to randomness. For example, at $N=20$, even a modest value of $\sigma_T \approx 31[m^{-1}]$ produces a nearly tenfold increase in the central peak, while at $\sigma_T \approx 50[m^{-1}]$ certain configurations yield enhancements of up to three orders of magnitude. These amplification peaks are not statistical artifacts but arise from a stochastic optimization mechanism: at low $\sigma_T,$ destructive interference dominates; as coherence breaks at intermediate $\sigma_T,$ contributions add incoherently, lifting the suppression; at high $\sigma_T,$ the broad delay distribution occasionally samples rare,favorable configurations where the relative delays accidentally align to produce strong constructive interference- more efficiently than in the regular or weakly random cases. Larger  pulse numbers increase both the combinatorial space for such optimal sequences and the sensitivity to delay variations, lowering 
the $\sigma_T$ threshold for observable enhancements.

Our results demonstrate that temporal randomness is not merely a source of spectral degradation but can - under specific conditions - be strategically exploited to enhance pair production.
 In particular, certain stochastic configurations strongly amplify the central peak of the distribution function, suggesting that tailored randomness in pulse sequences could serve as a resource for optimizing pair yields in strong-field QED. These findings open up new pathways for designing multipulse schemes in environments where perfect timing control is experimentally challenging.
 A systematic exploration of the optimal temporal arrangements that maximize pair creation, and a deeper analytical understanding of these rare enhanced configurations, remain compelling directions for future research.
 



%

\begin{figure}[tbp]\suppressfloats
	\centering
	\includegraphics[width=0.854\columnwidth]{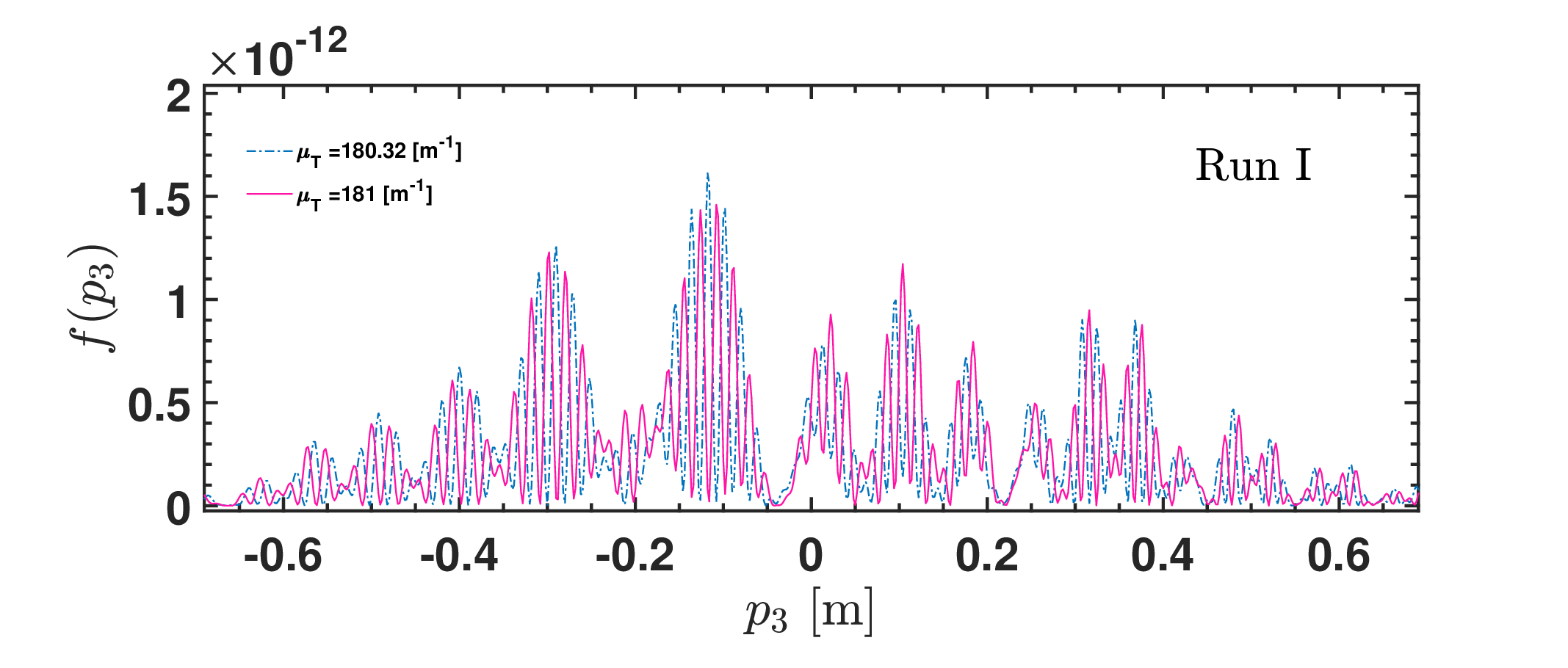}
	\includegraphics[width=0.854\columnwidth]{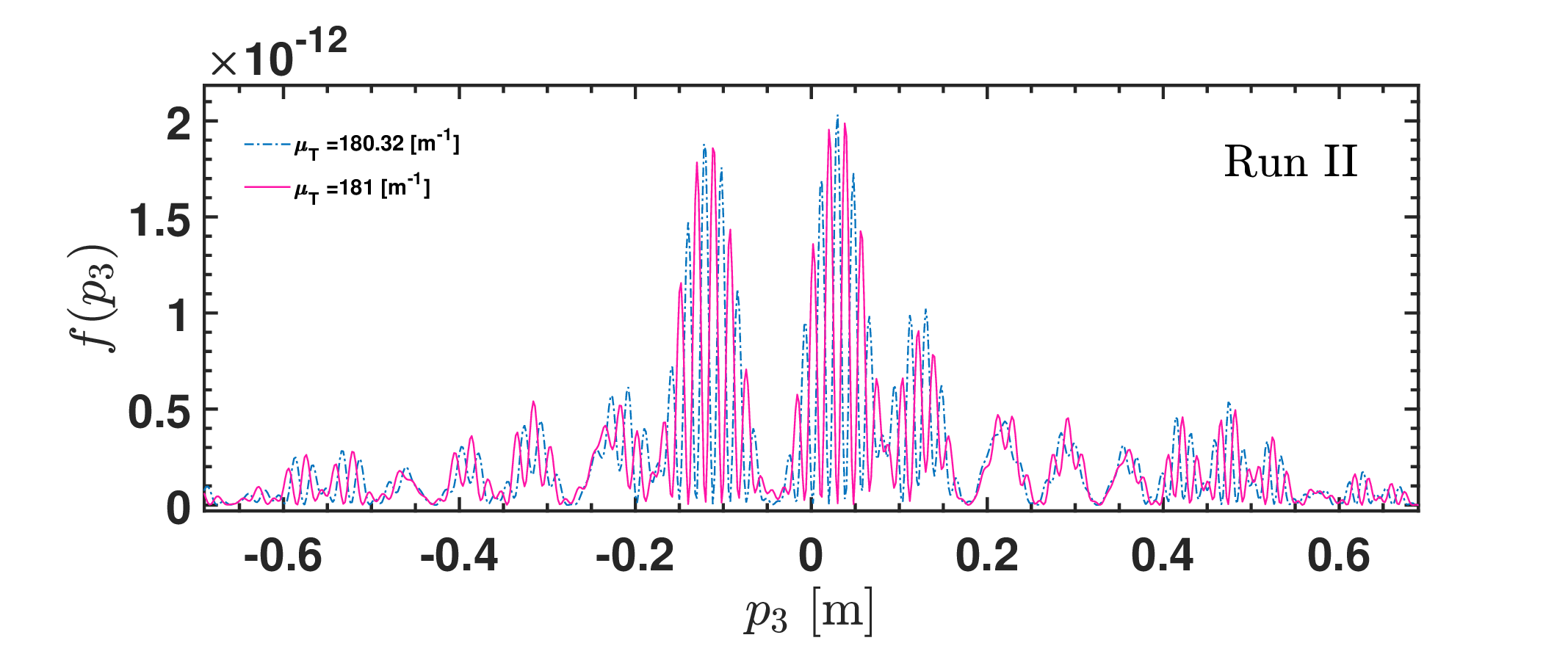}
	\includegraphics[width=0.854\columnwidth]{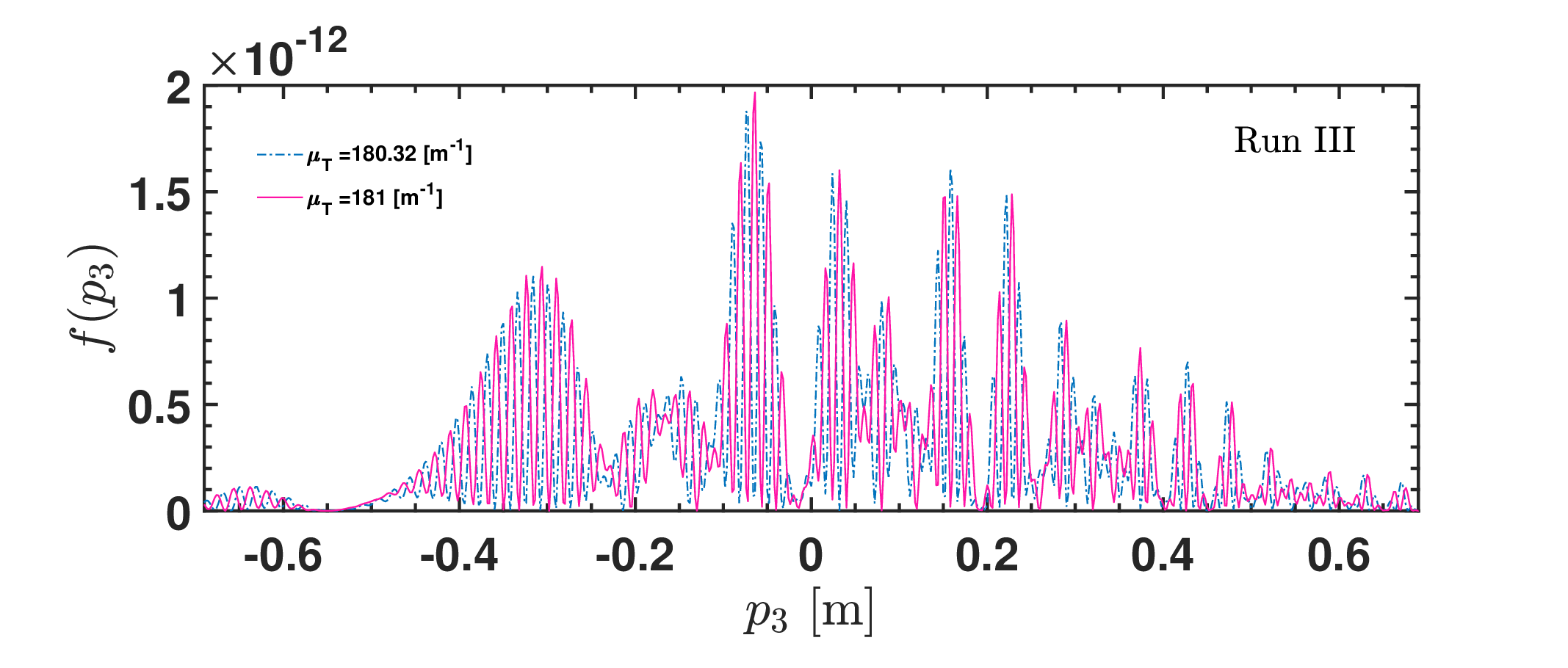}
	\caption{\label{181}Longitudinal momentum spectra for $N=4$ pulses with $\sigma_T=15~[m^{-1}]$. Parameters: $E_0=0.1E_c$, $\tau=20~[m^{-1}]$, $p_\perp=0$. (a) Run I: $\{T_k\} = \{172.93, 186.06, 161.54, 192.33\}\, [m^{-1}]$; (b) Run II: $\{T_k\} = \{194.01, 166.06, 194.76, 172.13\}\, [m^{-1}]$; (c) Run III: $\{T_k\} = \{195.64, 173.15, 183.65, 208.68\}\, [m^{-1}]$. All quantities are in electron mass units.}
\end{figure}
\section*{Acknowledgments}
We are grateful to the anonymous referee for constructive comments that helped improve the manuscript. Deepak Sah acknowledges financial support from the Raja Ramanna Centre for Advanced Technology (RRCAT) and the Homi Bhabha National Institute (HBNI).


\section{Supplementary }
\label{supply}

\subsection{Effect of  the mean inter-pulse delay $\mu_{T}$ on momentum spectrum}

\begin{figure}[tbp]\suppressfloats
	\centering
	\includegraphics[width=0.7754\columnwidth]{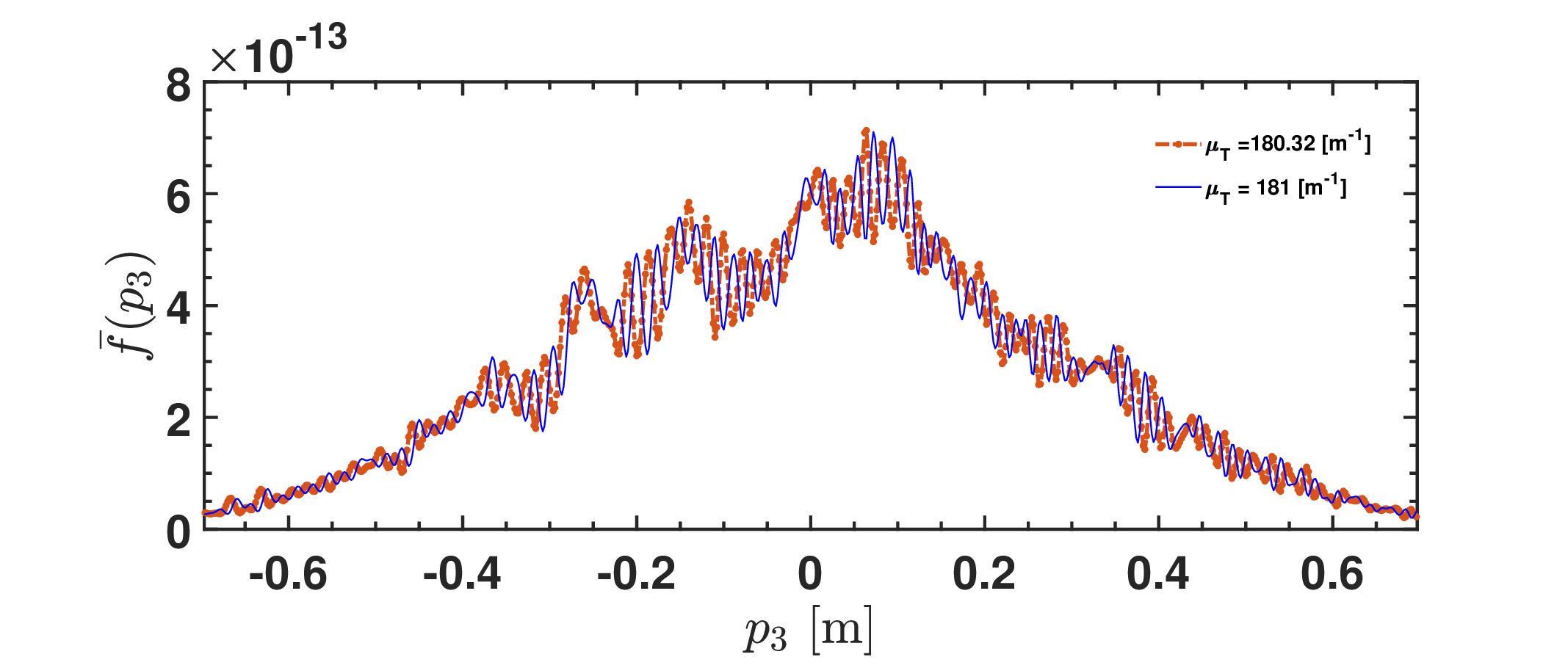}
	\caption{\label{181_avg}
	Averaged momentum spectra $\bar{f}(p_3)$ computed over 30-run of random samples with randomized time delays for an alternating-sign four-pulse electric field $E(t)$.  The field parameters are $E_0 = 0.1 E_c$, $\tau = 20 \,[m^{-1}]$, and $\sigma_T = 15 \,[m^{-1}]$.}
\end{figure}


The mean inter-pulse delay $\mu_T$ plays a decisive role in the coherent interference pattern of the momentum spectrum. In this supplementary section, we systematically analyze the effect of varying $\mu_T$ on both individual stochastic realizations and ensemble-averaged spectra, using $\sigma_T = 15 \; [m^{-1}]$ and $N = 4$ as a representative case.

In a regular ($\sigma_T = 0$) alternating-sign pulse train, $\mu_T$ acts as the temporal equivalent of the slit separation in optical multi-slit interference. The quantum phase difference accumulated between successive pulses scales proportionally to the product of the mean inter-pulse delay \(\mu_T\) and the longitudinal momentum \(p_3\).Therefore, even small changes in $\mu_T$ shift the condition for constructive and destructive interference, altering the entire fringe pattern.
Consequently, even minute variations  in $\mu_T$ shift the conditions for constructive and destructive interference, thereby reshaping the fringe pattern of the momentum spectrum.

Figure ~\ref{181} illustrates this sensitivity by comparing longitudinal momentum spectra for $\mu_T = 181\;[m^{-1}]$ (magenta) with the baseline value $\mu_T = 180.32\;[m^{-1}]$ (blue) across three independent stochastic realizations (Runs I–III). The mere $0.38\%$ increase in $\mu_T$ leads to discernible modifications in each realization. In Run~I, the central interference band shifts leftward, and the relative amplitudes of peaks are redistributed. In Run~II, the fringe spacing visibly changes, and side-band structures emerge at different momentum values. In Run~III, the symmetry about $p_3 = 0$ is broken, with one side of the spectrum noticeably enhanced relative to the other.

These observations confirm that, even in the presence of weak randomness ($\sigma_T > 0$), the interference pattern remains sensitive to $\mu_T$ on a realization-by-realization basis. This sensitivity stems from the fact that each stochastic configuration of inter-pulse delays interacts distinctively with the underlying mean temporal structure, thereby imprinting $\mu_T$-dependent phase shifts on the quantum interference.


Figure ~\ref{181_avg} shows the corresponding ensemble-averaged momentum spectrum computed over 30 stochastic realizations for $\mu_T = 181 \; [m^{-1}]$. Despite the sensitivity observed in individual runs, the averaged spectrum converges to a smooth, Gaussian-like envelope that is nearly identical to that obtained with the baseline value $\mu_T = 180.32 \; [m^{-1}]$. This demonstrates that, while individual stochastic realizations are $\mu_T$-sensitive, the ensemble-averaged result is robust. The underlying reason is statistical: averaging over many random delay configurations washes out the phase-sensitive interference details, leaving only the incoherent sum of contributions from individual pulses, which depends primarily on the pulse shape and amplitude, not on $\mu_T$.

The insensitivity of the averaged spectrum to small $\mu_T$ variations is encouraging for experimental applications. In realistic laser setups, where $\mu_T$ cannot be controlled with infinite precision, our results indicate several practical insights. First, single-shot measurements may show strong $\mu_T$-dependent variations due to the sensitivity of individual realizations. Second, however, multi-shot averaging will yield reproducible Gaussian-like spectra whose shape is largely independent of small systematic offsets in $\mu_T$. Third, the key parameter governing the transition from coherent to incoherent pair production is $\sigma_T$, not the exact value of $\mu_T$.

The value $\mu_T = 180.32 \; [m^{-1}]$ was chosen in the main text to facilitate direct comparison with Ref.~[46]. Our supplementary analysis shows that individual stochastic spectra are sensitive to $\mu_T$ due to phase-dependent interference, while ensemble-averaged spectra are robust against small $\mu_T$ variations. Consequently, the main conclusions of our work—regarding the role of $\sigma_T$ in driving the transition from interference-dominated to Gaussian-like spectra—remain valid across a range of $\mu_T$ values. Thus, while $\mu_T$ sets the coherent baseline, $\sigma_T$ controls the degree of decoherence, making our findings broadly applicable to experimental scenarios where timing cannot be perfectly regularized.

\bibliographystyle{elsarticle-num}

\bibliography{main} 

\end{document}